\pdfoutput=1 

\documentclass[journal=ansac3,manuscript=article, layout=twocolumn]{achemso}
\setkeys{acs}{keywords = true, articletitle = true}


\usepackage{placeins} 
\usepackage{bm} 

\usepackage{amssymb}

\usepackage{txfonts}

\let\oldbibitem\bibitem
\def\bibitem{\vfill\oldbibitem}

\usepackage[sort&compress,numbers,super]{natbib} 
\usepackage{achemso} 
\usepackage{todonotes}

\author{Gary W. Paterson}
\affiliation{SUPA, School of Physics and Astronomy, University of Glasgow, Glasgow, G12 8QQ, United Kingdom}
\email{Gary.Paterson@glasgow.ac.uk}

\author{Tsukasa Koyama}
\affiliation{Department of Materials Science, Osaka Prefecture University, Sakai, Osaka 599-8531, Japan}

\author{Misako Shinozaki}
\affiliation{Department of Basic Science, The University of Tokyo, Meguro, Tokyo 153-8902, Japan}

\author{Yusuke Masaki}
\affiliation{Department of Physics, The University of Tokyo, Bunkyo-ku, Tokyo 113-0033, Japan}

\author{Francisco. J. T. Goncalves}
\affiliation{Department of Physics and Electronics, Osaka Prefecture University, 1-2 Gakuencho, Sakai, Osaka 599-8570, Japan}
\alsoaffiliation{Chirality Research Center (CResCent), Hiroshima University, Higashi-Hiroshima, Hiroshima 739-8526, Japan}

\author{Yusuke Shimamoto}
\affiliation{Department of Physics and Electronics, Osaka Prefecture University, 1-2 Gakuencho, Sakai, Osaka 599-8570, Japan}

\author{Tadayuki Sogo}
\affiliation{Department of Physics and Electronics, Osaka Prefecture University, 1-2 Gakuencho, Sakai, Osaka 599-8570, Japan}

\author{Magnus Nord}
\affiliation{SUPA, School of Physics and Astronomy, University of Glasgow, Glasgow, G12 8QQ, United Kingdom}

\author{Yusuke Kousaka}
\affiliation{Research Institute for Interdisciplinary Science, Okayama University, Okayama, Okayama 700-8530, Japan}
\alsoaffiliation{Center for Chiral Science, Hiroshima University, Higashi-hiroshima, Hiroshima 739-8526, Japan}

\author{Yusuke Kato}
\affiliation{Department of Basic Science, The University of Tokyo, Meguro, Tokyo 153-8902, Japan}

\author{Stephen McVitie}
\affiliation{SUPA, School of Physics and Astronomy, University of Glasgow, Glasgow, G12 8QQ, United Kingdom}

\author{Yoshihiko Togawa}
\affiliation{Department of Physics and Electronics, Osaka Prefecture University, 1-2 Gakuencho, Sakai, Osaka 599-8570, Japan}
\alsoaffiliation{Chirality Research Center (CResCent), Hiroshima University, Higashi-Hiroshima, Hiroshima 739-8526, Japan}

\title[An \textsf{achemso} demo]
  {Order and Disorder in the Magnetisation of the Chiral Crystal CrNb$_3$S$_6$}

\keywords{Chiral soliton lattice, CrNb$_3$S$_6$, dislocation, differential phase contrast, transmission electron microscopy, Lorentz microscopy, Fresnel microscopy, ferromagnetic resonance spectroscopy}

\begin{document}
%
%

\begin{abstract}
Competing magnetic anisotropies in chiral crystals with Dzyaloshinskii Moriya exchange interactions can give rise to non-trivial chiral topological magnetisation configurations with new and interesting properties.
One such configuration is the magnetic soliton, where the moment continuously rotates about an axis.
This magnetic system can be considered to be one dimensional and, because of this, it supports a macroscale coherent magnetisation, giving rise to a tunable chiral soliton lattice (CSL) that is of potential use in a number of applications in nanomagnetism and spintronics.
In this work we characterise the transitions between the forced-ferromagnetic (F-FM) phase and the CSL one in CrNb$_3$S$_6$ using differential phase contrast imaging in a scanning transmission electron microscope, conventional Fresnel imaging, ferromagnetic resonance spectroscopy, and mean-field modelling.
We find that the formation and movement of dislocations mediate the formation of CSL and F-FM regions and that these strongly influence the highly hysteretic static and dynamic properties of the system.
Sample size and morphology can be used to tailor the properties of the system and, with the application of magnetic field, to locate and stabalise normally unstable dislocations and modify their dimensions and magnetic configurations in ways beyond that predicted to occur in uniform films.
\end{abstract}

\section*{Introduction}
Since the general theory of helicoidal magnetic structures was proposed~\cite{Dzyaloshinskii_JETP_helicoidal_1964} and experimentally confirmed in CrNb$_3$S$_6$,~\cite{togawa2012_fresnel_csl} there has been much interest in this~\cite{KISHINE2015_review, Togawa_jpsj_16_review} and other~\cite{Fert_2017_Nature_rev_skyrmion} chiral systems for the many unusual properties they possess.
The helimagnetism in CrNb$_3$S$_6$ originates from its monoaxial chiral hexagonal crystal belonging to the $P6_{3}22$ space group,\cite{Moriya_SSC_1982} which supports a large uniaxial anisotropy, $K_u$, along the chiral $c$-axis, symmetric Heisenberg exchange interactions, and anti-symmetric Dzyaloshinskii Moriya exchange interactions (DMI), $D$.~\cite{dzyaloshinsky1958, moriya1960}
The magnetisation in zero applied field ($H=0$) is in a chiral helimagnetic (CHM) phase, where the moment continuously rotates in the $ab$-plane, along the $c$-axis.
The intrinsic soliton periodicity in zero applied field, $L(H=0)$, depends on the ratio of the energy parallel to the $c$-axis, $J_{\parallel}$, to the DMI energy, and is 48~nm.~\cite{Moriya_SSC_1982, Miyadai_JPSJ_1983}

\begin{figure}[!hb]
  \centering
    \includegraphics[width=8.5cm]{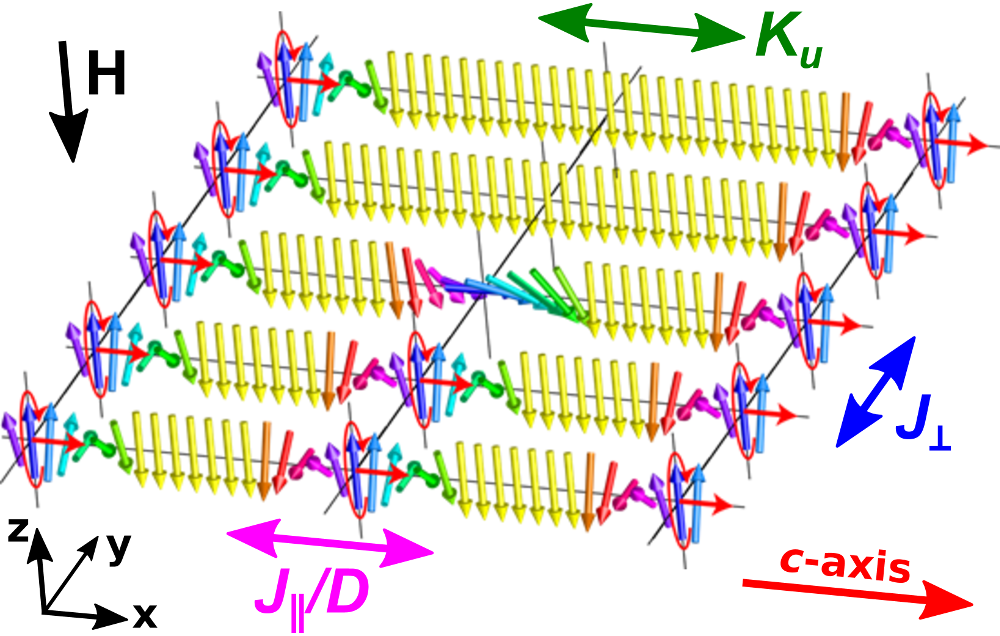}
    \caption{{Schematic representation of a chiral soliton lattice (CSL) with a dislocation created by a magnetic field, $H$, applied downwards. $J_{\perp}$ ($\sim 140$~K), $J_{\parallel}$ ($\sim 18$~K), $K_u$ ($\sim 1$~K), and $D$ ($\sim 2.9$~K) are the exchange energy density parallel and perpendicular to the $c$-axis (indicated by the red arrows), and the uniaxial anisotropy and DMI energy density, respectively.~\cite{Shinozaki_jpsj_2016}}
    \label{fig:schematic_dislocation}}
\end{figure}

When a magnetic field is applied perpendicular to the $c$-axis, a net-moment arises as the spins preferentially align with the magnetic field and the incommensurate CHM phase continuously transforms into the chiral soliton lattice (CSL) phase.
At sufficiently high fields, this creates regions of forced-ferromagnetic (F-FM) order, indicated by the yellow arrows in the schematic in Figure~\ref{fig:schematic_dislocation}, between which are isolated solitons where the moments rotate through 2$\pi$, as depicted by the orange to green coloured arrows.
As the field is increased, the F-FM regions grow in extent through distortion and then expulsion of solitons from the system until at some critical field, $H_c$, the entire system is in the F-FM phase.
Theoretical~\cite{Laliena_PRB_16_underHT} and experimental~\cite{Yonemura2017_PRB_phase_diac_oblique} phase diagrams for CrNb$_3$S$_6$ have been constructed.
The critical paramagnetic temperature depends on the crystal quality but is typically $\sim$127~K, while the critical field for CSL to F-FM transition varies due to sample thickness related demagnetisation effects, but is typically around 2~kOe.~\cite{Moriya_SSC_1982, Miyadai_JPSJ_1983}

Importantly, because the transition between the CSL and F-FM phases is continuous, the point at which the region between the magnetic kinks may be regarded as a F-FM \textit{domain} and not as part of a dilute CSL is not well defined, and will likely depend on the property of interest, \textit{e.g.} static or dynamic properties.
In this work, we take the phenomenological approach and refer to \textit{regions} of CSL or F-FM, depending on their most prominent character.

CrNb$_3$S$_6$ is unique amongst the helicoidal materials in that the magnetic ordering phase coherence extends over macroscopic dimensions.~\cite{togawa2012_fresnel_csl}
It is this protected property of the magnetic solitons, together with the ease with which the magnetic texture can be modified by the application of field, and the interesting emergent phenomena~\cite{KISHINE2015_review, Togawa_jpsj_16_review} that makes CrNb$_3$S$_6$ potentially useful in many device applications in nanomagnetism and spintronics.
Of particular interest are discretisation effects arising from changes in the magnetic topology created by the addition or removal of a soliton.
These effects have been observed in a variety measurement techniques, including transmission electron microscopy, magnetoresistance, magnetic torque, ferromagnetic resonance, and x-ray circular dichroism.~\cite{Yoshi_PRB_15_confinement, Yonemura2017_PRB_phase_diac_oblique, Francisco_PRB_17_FMR, Mito_PRB_18_geometrical_protection} 

Between the CSL and F-FM regimes, dislocations in the soliton lattice such as that depicted by the schematic in Figure~\ref{fig:schematic_dislocation} can form and are likely to play a key role in mediating the changes in the magnetic phase of the material.
Indeed, recently reported modelling predicts that soliton annihilation occurs through dislocations forming and moving perpendicular to the $c$-axis, while creation occurs in a coherent process at the sample edges.~\cite{Mito_PRB_18_geometrical_protection}
The coherence of the CSL is of great importance to the functionality of the material and it is thus crucial to characterise it.

In this work we experimentally characterise the transition from the F-FM phase to the CSL one and back in CrNb$_3$S$_6$ using Lorentz microscopy and ferromagnetic resonance spectroscopy, with a particular focus on magnetic lattice dislocations.
We show that dislocations mediate the CSL to F-FM transition.
In the opposite transition, dislocations appear as a result of the partitioning of the magnetisation into a mixture of CSL and field-polarised F-FM regions, creating a disordered phase.
The formation of this disordered phase depends on the reversal direction and creates an asymmetry in the associated dynamic properties.
The dislocations are generally metastable or unstable, but we show how they can be localised by `pinning' them against stepped increases in sample thickness, which also allows their dimensions to be modified far beyond that predicted from modelling of uniform thickness samples using a mean-field approach.
This observation opens the possibility to localise dislocations and make use of the nano-channels of field polarised spins created by them in device applications.

\section*{Results and Discussion}

\subsection*{Numerical Simulation of Chiral Soliton Lattice Dislocations}
We begin by describing the expected behaviour of edge dislocations formed in the chiral soliton lattice of a uniformly thick sample of CrNb$_3$S$_6$.
Here, we use the term `edge' to describe the dislocation in its crystallographic sense, not in the sample geometry one.
To do this, simulations were performed using a self-consistent mean-field approach to solving the spin Hamiltonian~\cite{Shinozaki_jpsj_2016} with the magnetisation formed from a three dimensional lattice of spins initialised with two dislocations.
A periodic system was defined with the $c$-axis along the $x$-axis, with 200 spins in $x$ and $y$, and 5 in $z$ (see Figure~\ref{fig:schematic_dislocation} for orientation definitions).
The magnetisation parameters used were $D/J_\parallel = 0.16$, $J_\perp / J_\parallel =
8$, $H_z / (J_\parallel S) = 0.01$, $H_\parallel / (J_\parallel S) = 10^{-14}$, and $T / (J_\parallel S^2)=1$, where $S$ and $T$ are the spin modulus, and the temperature, and $J_\parallel$ and $J_\perp$ are the exchange energy densities along the $x$-axis (parallel to the $c$-axis) and perpendicular to it, respectively (see Methods for further details).

\begin{figure*}[!htb]
  \centering
    \includegraphics[width=17.5cm]{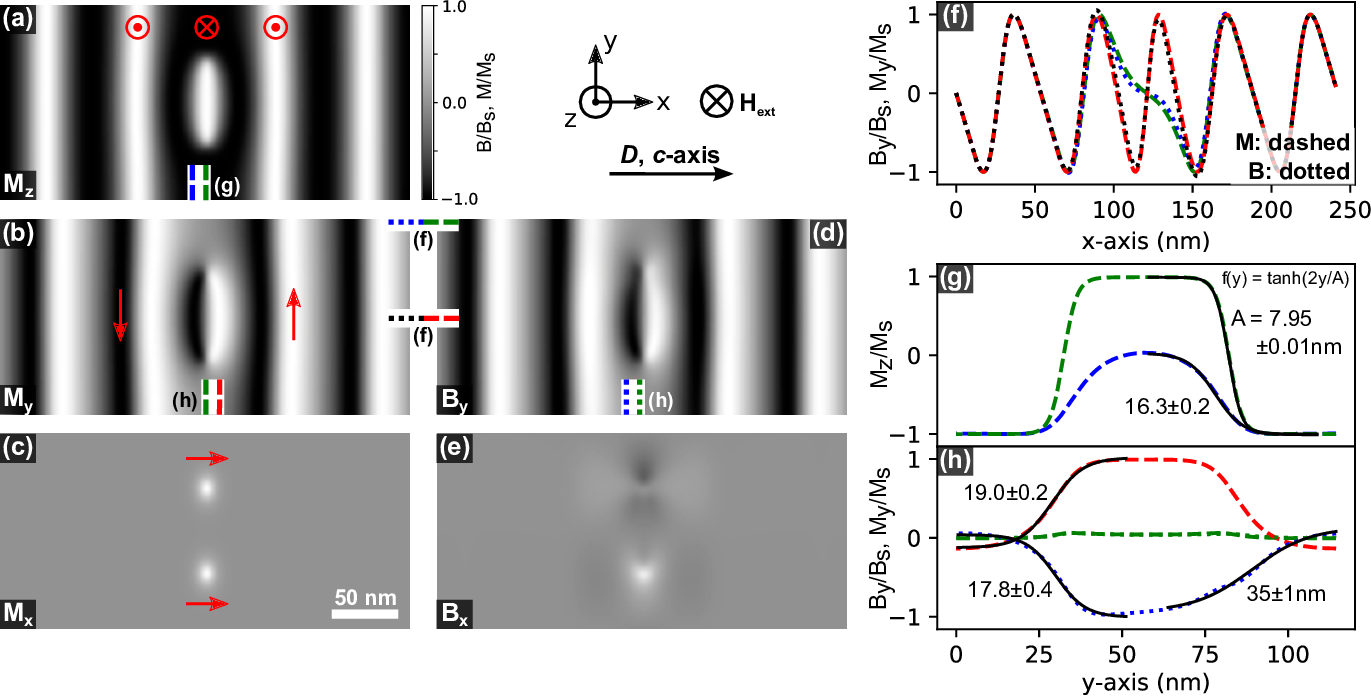}
    \caption{{CSL dislocation simulations with the field applied perpendicular to the $c$-axis. The left hand column (a-c) shows the projected magnetisation components in an untilted sample. The red arrows indicate the magnetisation direction. The middle column (d-e) are the calculated integrated magnetic induction components that would be imaged by DPC measurements. The right hand column (f-h) are the profiles at the locations indicated by the coloured lines in the surface plots. Dashed lines are $M$, dotted lines are $B$. The annotated dimensions are widths of hyperbolic tangent functions (black lines) fitted to the data, as defined in the figure. All data are presented in normalised units and with a common colour map.}
    \label{fig:numerical_simulations}}
\end{figure*}

The magnetisation distribution does not converge upon iteration of the calculations, indicating the dislocations are unstable and, instead, the dislocations approach one another and eventually annihilate.
In doing so, the spins forming the soliton `untwist' to lie parallel to one another in the direction of the field, creating a region of more dilute CSL.
The projected magnetisation components part way through this process are depicted in the left hand column of Figure~\ref{fig:numerical_simulations} and are characteristic of the general distribution.
Oscillations in intensity along the $x$-axis in panels (a) and (b) represent rotation of the moment in the $ab$-plane along the $c$-axis, with 2$\pi$ rotations of the moment between successive peaks or troughs.
At the positions of the dislocations, the moment is predicted to rotate in-plane to point along the $c$-axis, creating a non-zero $x$-component of magnetisation as shown in Figure~\ref{fig:numerical_simulations}(c), similar to that depicted in Figure~\ref{fig:schematic_dislocation}.

To enable comparisons with differential phase contrast (DPC) measurements (discussed later), we display in the middle column of Figure~\ref{fig:numerical_simulations} the calculated integrated induction field components in the plane of the sample that would be imaged in the experiment, using the Fourier space approach to solving the Aharonov-Bohm equation.\cite{Beleggia_apl_2003_mag_sim}
With the electron beam travelling along a path perpendicular to the film, DPC is insensitive to the $z$-component of the magnetisation.
The projected in-plane magnetisation and induction will differ if the magnetisation has any divergence.~\cite{Stephen_JAP_2001}
This is true for the dislocations, so it is important to understand the differences between the magnetisation from the model and the measured induction field from Lorentz microscopy.
The magnetisation divergence is plotted in Figure~\ref{fig:simulation_m_div_curl}, and is accompanied by a detailed discussion.

The right hand column (f-h) of Figure~\ref{fig:numerical_simulations} shows profiles along the lines in the surface plots (a, b, d) along with fits to the data (thin black lines in (g) and (h)) to estimate the feature widths.
The $y$-component of the magnetisation (b) shows the anti-parallel configuration of the CSL and this is well mapped by the induction field (d) at most locations, as indicated by the overlapping data in panel (f), with subtle differences in the vicinity of the dislocations where there is some divergence of the magnetisation.
However, the $x$-component of the magnetisation (c) results from the presence of the dislocation and is divergent in nature. 
Divergence of magnetisation results here from both the $x$-and $y$-components and results in magnetic field sources ($H$) which will make the projected magnetisation and induction components differ.
In this case, the non-divergent magnetisation dominates the contrast in (d) which is only slightly different from (b), whilst in (c) the magnetisation is mostly divergent and so, not unexpectedly, the contrast in (e) is quite different from in (c).
It can be seen that the contrast in (e) is significantly reduced compared to (c) and that opposite ends of the dislocation have distinctly different character.
These differences are shown in more detail in the difference and histogram plots in Figures~\ref{fig:numerical_simulation_HM_error} and \ref{fig:numerical_simulation_histograms}, respectively.

For the magnetic configuration depicted in Figure~\ref{fig:numerical_simulations}, the moment continually rotates between solitons (see green and blue lines in (f)), showing that the system may be regarded as a dilute CSL.
However, in both the F-FM \textit{and} CSL phases, the spins in the $ab$-plane (the $yz$-plane of Figure~\ref{fig:schematic_dislocation}) are parallel to all other spins within the same plane.
A dislocation in a CSL marks the location where the spins in the $ab$-plane rotate between regions of different spin orientations, as shown in Figure~\ref{fig:numerical_simulations} and, thus, they may be considered as a kind of domain wall.
As in normal ferromagnets, the domain wall width here is predicted to be independent of the applied field strength, and to be proportional to $\sqrt{J_\perp/K_{eff}}$, where $J_\perp$ is the exchange energy density in the $yz$-plane, perpendicular to the $c$-axis, and $K_{eff}$ is the effective anisotropy which is determined by $J_\parallel$, and the DM interaction ($D$) and uniaxial anisotropy ($K_u$) strengths: $K_{eff} = 2\left(\sqrt{J_\parallel^2+D^2} - J_\parallel \right) + K_u$.~\cite{Kishine_PRB_2009_spin_resonance}

The wall width in this system is most naturally defined as the distance over which the $z$-component of the magnetisation reverses.
To extract this width, we fit hyperbolic tangent functions to the data, as shown in Figure~\ref{fig:numerical_simulations}(g), and find that the best fitting parameters for the profiles are 7.95~nm down the centre of the soliton dislocation and 19.0~nm either side of this.
Experimentally, we do not measure this component with DPC imaging, however it gives useful lengthscales.
Therefore, to be able to compare with experiment, the $y$-component of the integrated induction at its position of maximum (or minimum) strength, displaced laterally from the soliton centre is used as a measure of the dislocation width. 
By comparison of Figures~\ref{fig:numerical_simulations}(a) and (b), it is clear that this width will be much wider than the distance over which the $z$-component of the dislocation magnetisation, $M_z$, reverses.
Line profiles of the $y$-components of the magnetisation and induction field at the positions indicated by the dotted and dashed lines, respectively, in the images in Figure~\ref{fig:numerical_simulations}(b) and (d) are plotted in Figure~\ref{fig:numerical_simulations}(h) and show that a width parameter of either 17.8 or 35~nm is obtained from the induction variation. 
At one end of the dislocation, the effect of the $H$ field due to the magnetisation divergence is low and the induction and magnetisation profiles are very similar, with a width profile of 17.8~nm. 
However, at the other end there is a much larger difference due to the increased $H$ field created from the magnetisation divergence and the width is larger at 35~nm (see supplemental information for a detailed discussion).
In terms of the measured numbers, we see that the smaller number of 17.8~nm is comparable to the $y$- and $z$-components of the magnetisation off centre of the dislocation.

\subsection*{Experimental Observations of Dislocations}

\begin{figure*}[!hbp]
  \centering
    \includegraphics[width=14.0cm]{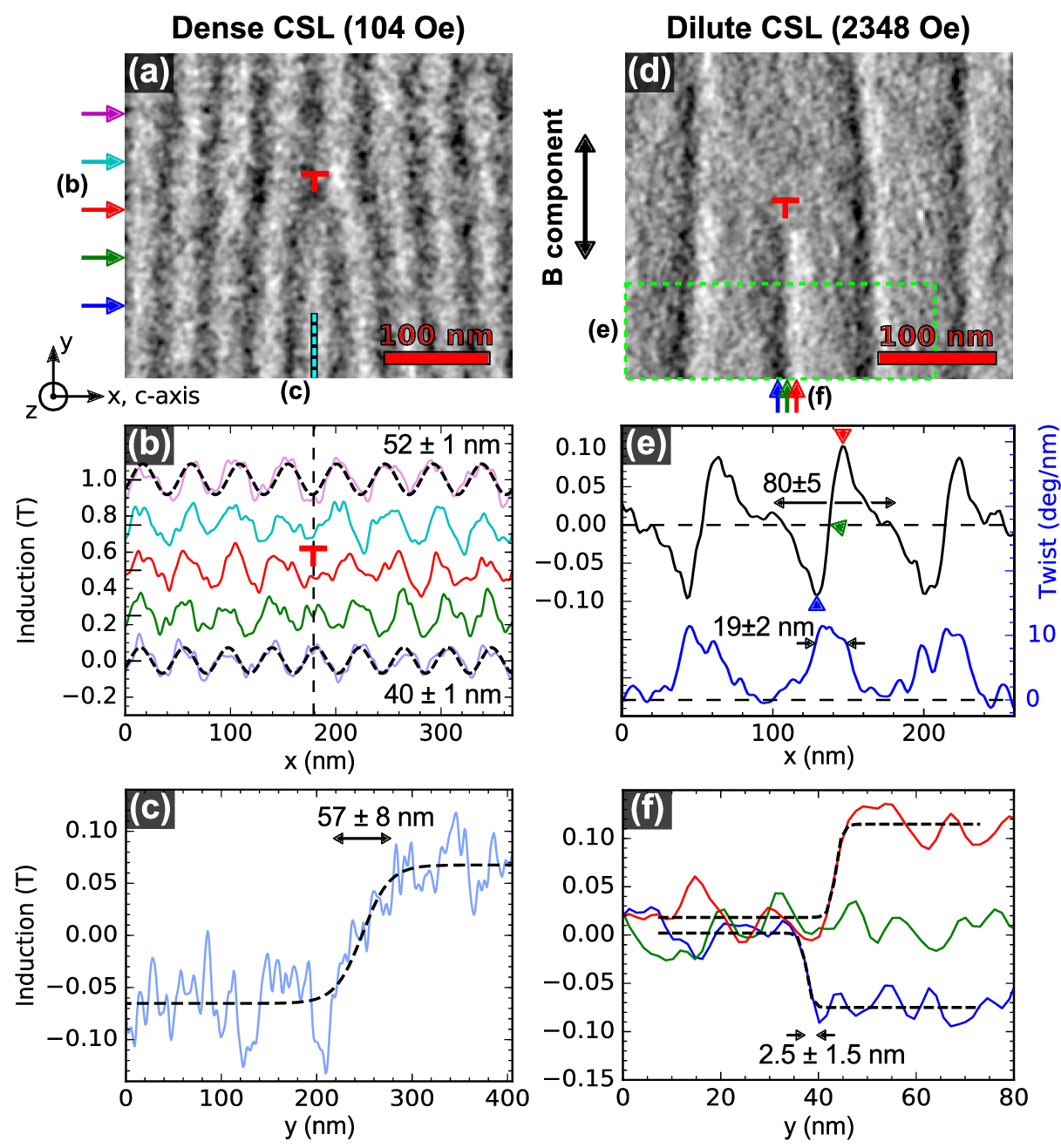}   
    \caption{{Induction maps of dislocations at low (left column, 104~Oe) and high (right column, 2348~Oe) fields, measured using DPC imaging at a temperature of 102~K. (a) and (d) show the component of the induction perpendicular to the field applied into the page, where the dislocations are marked by a red `$\perp$' symbol. (b) Line profiles along the $c$-axis at positions indicated by the arrows in (a). (c) Induction profile along the centre of the low field dislocation, as marked by the vertical dashed line in (b). Soliton profiles parallel and perpendicular to the $c$-axis at high field are shown in (e) and (f), respectively. The coloured arrows in (d) and (e) indicate common locations, and are those of the matching coloured lines in (f). The data in (e) was produced from average of the region outlined with a green dashed line in (d). The dashed lines in (b), (c) and (f) are fits to the data of sinusoidal and hyperbolic tangent functions. The lower (blue) line in (e) is the rotation rate of the high field solitons. See Methods for analysis details.}
    \label{fig:dpc_dislocation}}
\end{figure*}

To directly observe edge dislocations formed in the chiral soliton lattice of CrNb$_3$S$_6$, the magnetisation state of a thin ($\le$80~nm thick) lammela sample was investigated using DPC imaging in a scanning transmission electron microscope (see Methods for details of the sample preparation and imaging).
Unlike Fresnel imaging (discussed later), DPC is an in-focus Lorentz imaging technique~\cite{CHAPMAN1999729} and, with a known sample thickness, allows us to quantitatively measure with high spatial resolution the sample magnetic induction, from which the magnetisation state can be inferred.
The dislocations, marked in Figure~\ref{fig:dpc_dislocation} by a red `\rotatebox[origin=c]{180}{$\perp$}' symbol, are shown with a low (left column) and high (right column) field applied perpendicular to the $c$-axis.
The measured $x$-components of induction (along the $c$-axis) are shown in Figure~\ref{fig:induction_components}, and will be discussed later.
Unlike in the simulations, in the experiment, these dislocations are metastable at low fields and can be made stable at high fields by sample thickness modulation.

In the low field case, the stability of the dislocation may not be captured in the simulation due to the finite size of the sample and periodic boundary conditions being imposed, which limits the extent to which the soliton lattice can distort.
Instead, the simulation may be more representative of a dilute CSL phase.
In the real system, the soliton lattice is dense at low applied field values and gradual lattice deformation occurs around the dislocation, as may be seen in Figure~\ref{fig:dpc_dislocation}(a).
In this case, the equivalence of dislocations and domain walls is somewhat modified; instead of defining a wall between regions of CSL and F-FM, the dislocation may be thought of as defining regions of different soliton kink densities.
The large lateral size of the experimental sample means that there are very many solitons and addition or removal of one soliton can be accommodated more easily through gradual deformation of the lattice.

Induction profiles parallel to the $c$-axis at the positions of the arrows in Figure~\ref{fig:dpc_dislocation}(a) are plotted in panel (b), and the period of the top and bottom series estimated by fitting sinusoidal functions to the data.
We use sinusoidal functions since the field used here (104~Oe) is over an order of magnitude smaller than the typical critical field, so the magnetisation closely resembles that of the helimagnetic phase expected from the Sine-Gordon model,~\cite{Kishine_PTPS_05} the projection of the moment of which is sinusoidal.~\cite{togawa2012_fresnel_csl} 
The CSL periodicity increases from 40~nm below the dislocation to 52~nm above it, straddling the expected low temperature CHM value of 48~nm, and reflects local distortion of the lattice in the vicinity of the dislocation, as discussed above.
From the curvature of the solitons in Figure~\ref{fig:dpc_dislocation}(a), clearly the lattice deformation is strongest near the dislocation and it extends over multiple soliton periods in all directions.
The local periodicity extracted from this data is shown in supplemental Figure~\ref{fig:dislo_period}(b) and confirms distortion extends over 100~nm in directions along and perpendicular to the $c$-axis.

The measured induction profile perpendicular to the $c$-axis, along the centre of the dislocation is shown in Figure~\ref{fig:dpc_dislocation}(c).
The dislocation is approximately 60~nm wide and is much narrower than the long range lattice distortion, reflecting the relatively weak DMI strength compared to that of the uniaxial anisotropy.
This can be understood by considering that the magnetisation rotation rate reduces in the immediate vicinity of the dislocation, increasing the DMI contribution to the energy, yet the moment remains very closely aligned with $ab$-plane throughout because of the larger uniaxial anisotropy contribution to the energy of moments pointing away from the easy-axis.

Equivalent induction plots for the high field case in which the CSL is diluted by wide regions of slowly rotating moments are shown in the right hand column of Figure~\ref{fig:dpc_dislocation}.
At high fields, dislocations are generally unstable in uniform thickness films, consistent with the simulations, but can, as done here, be stabilised by pinning them up against an energy barrier created by an increasing sample thickness.
In the sample images presented here, the thickness begins to increase to several hundred nanometres from just below the bottom edge of the image in Figure~\ref{fig:dpc_dislocation}(d) (see Figure~\ref{fig:sample_overview} for further details).

The soliton profile parallel to the $c$-axis (black-line) and the net moment twist rate (blue line) produced from the average of the data outlined with a green dashed line in (d) are shown in Figure~\ref{fig:dpc_dislocation}(e).
Here, the blue, green and red arrows indicate equivalent points to those similarly marked in Figure~\ref{fig:dpc_dislocation}(d) and the locations of profiles drawn in the same-coloured lines in Figure~\ref{fig:dpc_dislocation}(f) (discussed later).
Since the magnetisation at locations far from dislocations is free of divergence, the saturation magnetisation can be inferred directly from the integrated induction from the DPC measurement and is $\mu_o M_s \approx 0.07-0.1$~T.
This is similar to values reported from selected area electron diffraction measurements of lamella at 110~K of 0.086~T, corresponding to 1.5~$\mu_b$/Cr,~\cite{togawa2012_fresnel_csl} and of 1.9~$\mu_b$/Cr~\cite{Miyadai_JPSJ_1983} in bulk samples at a temperature of 102~K.
The average trough to peak (blue to red arrow) distance is $\sim$19~nm, and is considerably narrower than half the $\sim80$~nm full-width of the soliton, shown in Figure~\ref{fig:dpc_dislocation}(e), demonstrating how the soliton structure is modified at high fields to reduce the Zeeman energy.
This can be seen in panel (e) by the asymmetry of the induction gradient on each side of the maxima or minima, matching that in the simulation (\textit{c.f.} Figure~\ref{fig:numerical_simulations}(f)), and appears as a narrower rotation between troughs and peaks in the twist rate data when moving from left to right along the graphs (between the blue and red arrows).
The maximum rate is $\sim11^\circ$/nm, which would give a period of only $\sim$33~nm if the rate is maintained across the soliton.
The region of faster rotation will correspond to that in the sample where a component of the moment points anti-parallel to the applied field which, in this case, points into the page.

The three coloured lines in Figure~\ref{fig:dpc_dislocation}(f) show the induction profile across the dislocation, perpendicular to the $c$-axis, at the points indicated by the arrows of the same colour in panels (d) and (e). 
The blue and red traces are those at the minima and maxima, respectively, and show that the dislocation width is $\sim$2.5~nm.
We note that while the dislocation did not change appearance prior to, during and after the measurement, we cannot rule out the end of the dislocation being dynamic and moving on a time frame below the exposure time, blurring the features.

In addition to the dislocation stability, compared to the simulations, there are two key differences in the experiment.
First, the non-zero $x$-component of induction is not observed in the experiment at low or high fields, as shown in Figure~\ref{fig:induction_components}.
In both cases, this may be in part because of the reduced sensitivity of the imaging technique to this particular component, as discussed above (see Figure~\ref{fig:numerical_simulation_HM_error}) and, potentially, because the dislocation may be dynamic (see Figure~\ref{fig:sup_dpc_slow_dynamics}, discussed later).
At high fields, the dislocation may additionally be modified by being pinned against a barrier.
It is, however, unclear why no $x$-component is seen at low field, but it may be related to the dense nature of the lattice in the experiment which would favour the moments remaining aligned with the $ab$-plane.
Secondly, the dislocation widths along the $ab$-plane can be significantly modulated by the applied field strength, from approximately 60~nm at low field to around 2.5~nm at high field.
Since these measurements are of the $y$-component of the induction, the z-component of the magnetisation reversal width is very likely to be much smaller.
These deviations from the simulations open the possibility to reproducibly locate and stabilise normally unstable or metastable dislocations and to modify their properties in ways not possible in uniform thickness films.

\begin{figure*}[!htb]
  \centering
    \includegraphics[width=12.0cm]{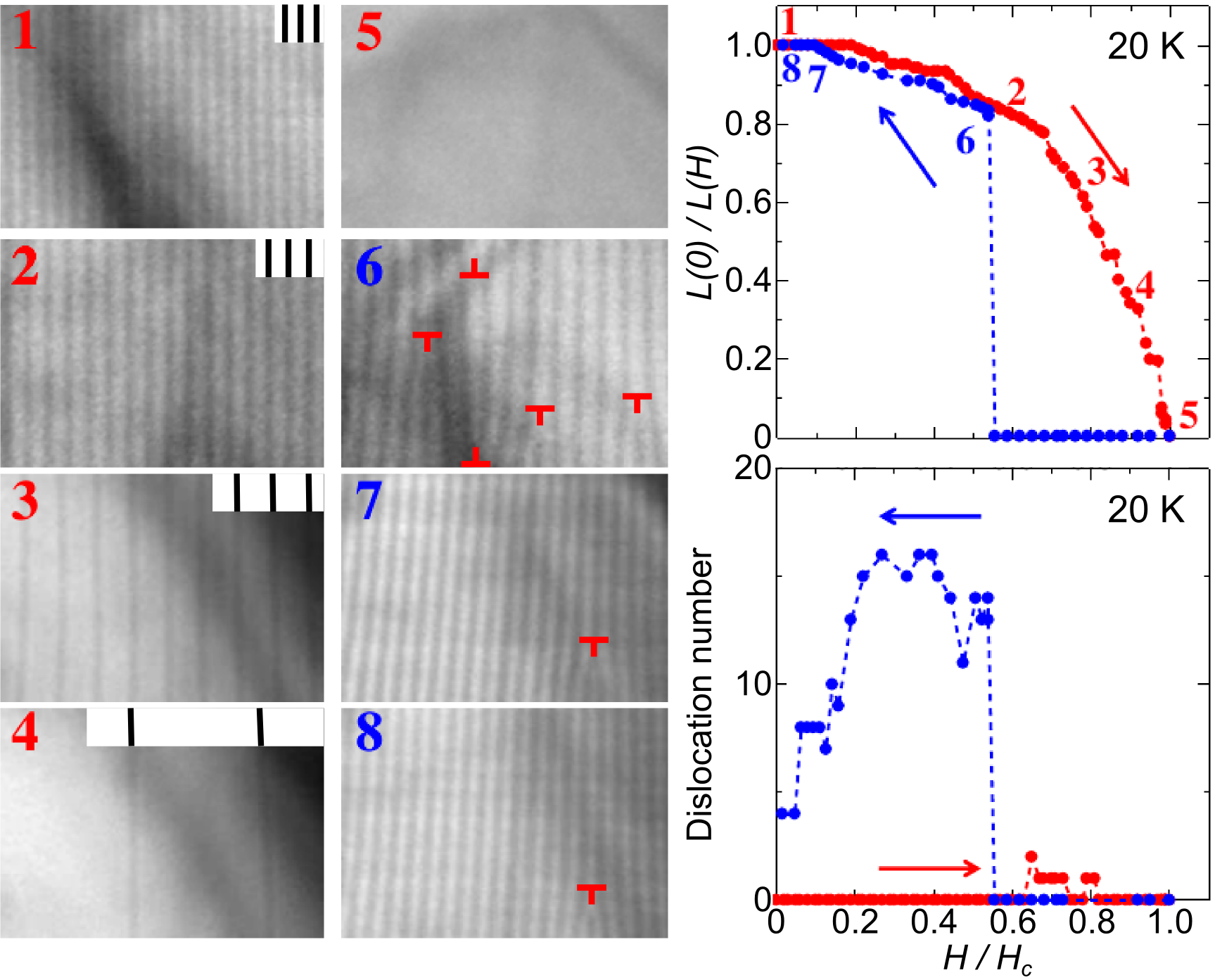}   
    \caption{{Hysteretic field dependent soliton density and dislocation population in a  CSL, obtained from Fresnel imaging of a thin-film lamella at 20~K with the field, applied perpendicular the $c$-axis and to the sample surface, swept in both the increasing (red labels) and decreasing (blue labels) directions, as indicated by the arrows in the right hand panels. 960$\times$670~nm sections of the Fresnel images are shown in the left hand panels, where dislocations are marked by red $\perp$ symbols, while statistics were gathered from a fixed area of 2060$\times$1540~nm.}
    \label{fig:fresnel_dislocation_populations}}
\end{figure*}

\subsection*{Hysteretic Chiral Soliton Lattice Dislocation Populations}
The formation and subsequent movement of dislocations is one means by which solitons may be nucleated or annihilated~\cite{Mito_PRB_18_geometrical_protection} and thus we may expect them to play a significant role in transitions to and from the F-FM and CSL phases.
To investigate this, we observed the transition between these phases by Fresnel imaging~\cite{CHAPMAN1999729} of a thin lammela sample of several microns in size.
The results of this are depicted in Figure~\ref{fig:fresnel_dislocation_populations} and show that the dislocation properties strongly depend on whether the transition is from F-FM to CSL or \textit{vice versa}.
Solitons in the Fresnel images in the left hand panels appear at high fields as thin dark vertical lines on a variable intensity background (occurring due to bend contours in the weakly curved sample diffracting intensity away from the detector by different amounts), as shown by the black and white overlays in the left hand column.
Panels 1-5, labelled with a red font, show the transition from the CSL phase to the F-FM one as the applied field strength increases from zero to the critical field, $H_c$, while panels 6-8 (blue font) show the return to zero field.
The results of analysis of this and data over an area five times wider than that shown in the figure at intermediate fields to extract the field dependence of the soliton period, $L(H)$, and dislocation density are shown in the upper and lower right hand panels of Figure~\ref{fig:fresnel_dislocation_populations}, respectively.

The CSL to F-FM transition (red symbols in Figure~\ref{fig:fresnel_dislocation_populations}) is characterised at low fields by a relatively smooth, near-continuous stepped increase in soliton period as solitons are removed from the system in a stochastic process.
At higher fields, there are fewer solitons and removal of a single soliton has a proportionally larger effect, resulting in increasingly large jumps in period.
Importantly, very few dislocations form across the entire phase transition.
This suggests that expulsion of solitons from the system is a coherent process that maintains the long range lattice order, and that the energy barrier to annihilation is relatively low.
These observations are consistent with reports of annihilation occurring due to `unzipping' of solitons as the result of formation and rapid movement of a dislocation perpendicular to the $c$-axis,~\cite{dussaux_ncoms_16_FeGe_dynamics, Mito_PRB_18_geometrical_protection} and with our observations of dislocation movement, discussed later.

The F-FM to CSL transition (blue symbols in Figure~\ref{fig:fresnel_dislocation_populations}) shows markedly different behaviour; the system remains in the F-FM phase until the field is reduced to around 0.55~$H_c$ (labelled point 6), at which field a high number of solitons appear as the phase changes from a `supercooled' F-FM phase to a CSL one with a similar periodicity to that at the same field in the opposite transition direction, but with very many dislocations.
The number of dislocations remains relatively high throughout the transition to the CSL phase at a near-zero field, gradually reducing in number as the field is decreased and metastable dislocations are removed and new solitons enter the lattice.

We note that a few solitons remain at near-zero field in the decreasing branch, indicated by point 8 in Figure~\ref{fig:fresnel_dislocation_populations}.
The applied field strength at this point was set by the residual field of the transmission electron microscope objective lens and was around 100~Oe.
However, when the field polarity is swapped, equivalent to point 1 in Figure~\ref{fig:fresnel_dislocation_populations}, the dislocations disappear and the ideal CSL is obtained.
This picture is consistent with the behaviour observed in ferromagnetic resonance (FMR) experiments (discussed later) when sweeping the field applied to a sample in the F-FM state through zero.

Our direct observations via Lorentz TEM (at a lower temperature 20~K) strongly supports the picture obtained based on the magnetoresistance data already reported in samples with similar dimensions.~\cite{Yoshi_PRB_15_confinement}
In smaller volume samples, different results are obtained. 
When the sample is reduced in size by crystallographic chirality,~\cite{Yoshi_PRB_15_confinement} soliton dislocations appear at the edge of the sample and move inwards, towards a step increase in thickness, as shown in Figure~\ref{fig:nucleation_in_grain}.
Because of the small volume and the stabilising effect of the thickness step, the number of nucleation sites is reduced and more symmetric behaviour with respect to the field cycle is seen in the dislocation densities.

This sample size dependent hysteretic behaviour is similar to that observed in magnetoresistance and magnetic torque measurements on micron-sized samples~\cite{Yoshi_PRB_15_confinement, Yonemura2017_PRB_phase_diac_oblique} and suggests there is a significant energy barrier to initial soliton nucleation.
This is consistent with recent modelling efforts which predict that the transition from a supercooled F-FM to CSL phase occurs in small samples by the creation and penetration of solitons from the edge of the sample~\cite{Mito_PRB_18_geometrical_protection}, and that a barrier of magnitude $(4/ \pi^2 \approx 0.4) H_c$ is formed by twists of the magnetisation at the surface of the sample from which they enter.\cite{Shinozaki_PRB18_soliton_barrier}
We note this value is similar to the one we observe in the TEM data in Figure~\ref{fig:fresnel_dislocation_populations} of 0.55~$H_c$, and that the precise value will vary with sample-dependent demagnetisation effects.~\cite{Shinozaki_PRB18_soliton_barrier}

In all but the smallest samples, the process of solitons entering the system may occur from several edges or nucleation points, depending on the local demagnetisation field~\cite{FTG_PRB18_tailored_FMR} and, where the CSL lattices meet, metastable dislocations may form.
Evidence of dislocation instability is shown in supplemental Figure~\ref{fig:sup_dpc_slow_dynamics}, which demonstrates that dislocations in CrNb$_3$S$_6$ at constant temperature and under a constant applied field exhibit slow dynamics, on the order of seconds, similar to observations of dislocations in the helimagnetism of FeGe.~\cite{dussaux_ncoms_16_FeGe_dynamics}

\subsection*{Resonance Properties of the Chiral Soliton Lattice}

\begin{figure}[!htb]
  \centering
    \includegraphics[width=8.5cm]{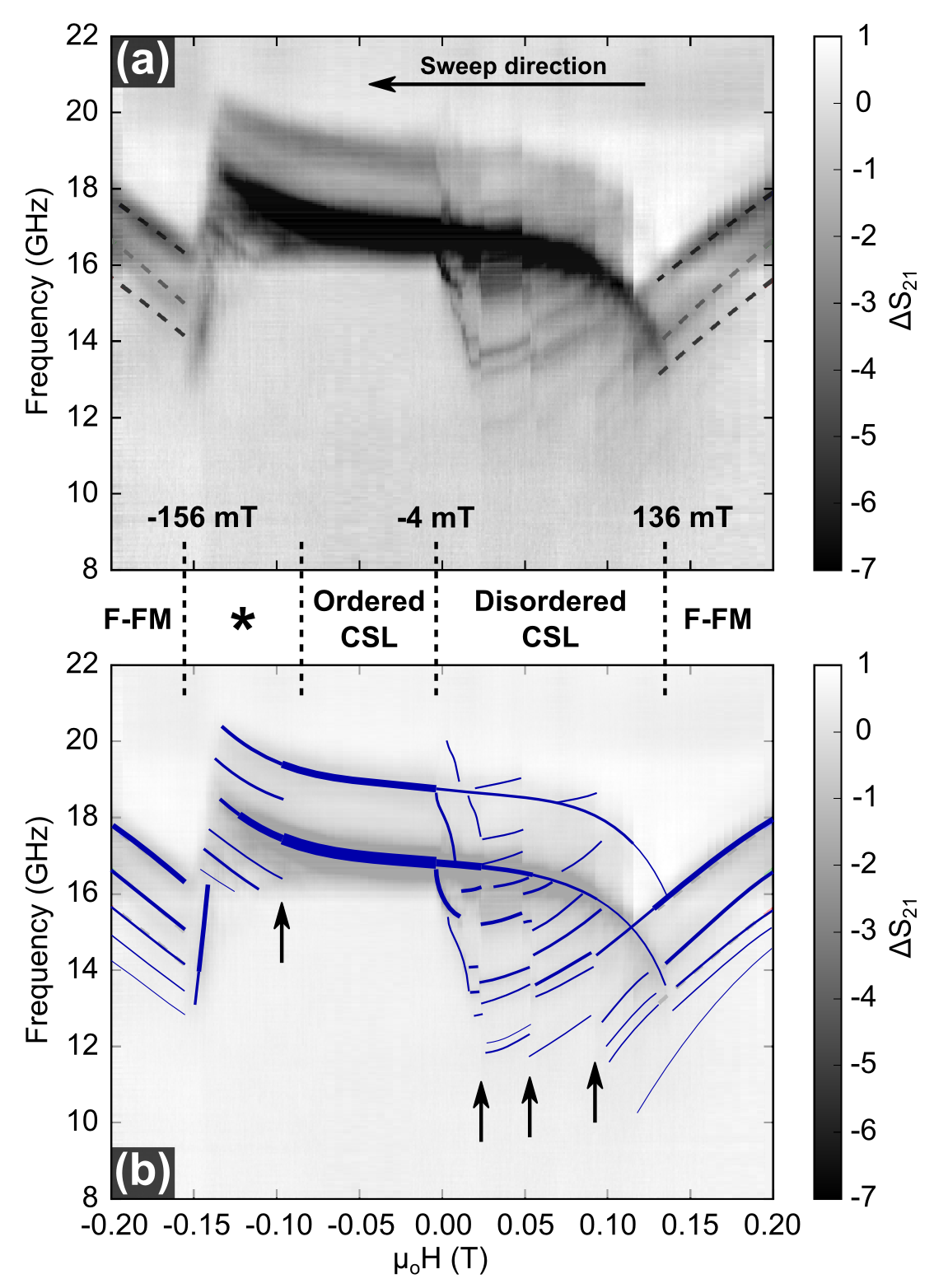}
    \caption{{(a) Ferromagnetic resonance measurements on a specimen with a 50~$\muup$m long chiral axis at 50~K as the DC applied field is varied from +0.2~T to -0.2~T in 1~mT steps, showing how the dynamic magnetic properties are influenced by the magnetic phase. (b) A simplified sketch of (a) where the lines pick out the main resonance branches. The vertical arrows highlight examples where multiple branches jump in frequency together.}
    \label{fig:fmr}}
\end{figure}

In addition to dislocations exhibiting slow dynamics, one might expect them to also influence the fast magnetisation dynamics of the system.
However, since the volume and coherence of the dislocations will be relatively small, the direct effect will most likely be limited to modification of the resonance linewidth, as is predicted to occur in non-chiral ferromagnets with crystal defects.~\cite{Baryakhtar_JETP_68_FM_dislocation}
Instead, we expect that the dominant effect of dislocations will be the indirect one, where the presence or movement of dislocations may either mediate the creation or indicate the presence of multiple CSL and F-FM regions within the sample, depending on the sweep direction, as discussed in the previous section.
In addition, dislocations may also define multiple regions of dense lattices between which exist relative periodicity and phase changes in the rotation of the moment.

To probe the microwave frequency dynamics of the material in its different phases, we performed ferromagnetic resonance (FMR) measurements on a sample with a 50~$\muup$m long chiral axis at 50~K in a custom system~\cite{Francisco_PRB_17_FMR} comprised by a vector network analyser coupled to a broadband waveguide (see Figure~\ref{fig:fmr_sup} for further details).
FMR data recorded while the field was swept from +0.2~T to -0.2~T, after normalisation and processing (see Methods for details), is shown in Figure~\ref{fig:fmr}(a).
To aid interpretation of the complex spectra, the main resonance branches are highlighted in the sketch overlay shown in Figure~\ref{fig:fmr}(b).
In the following, we will restrict our discussion to the main features of the FMR data.
A more detailed FMR study will be reported elsewhere.

The general FMR features are similar to those reported previously from a smaller sample with with the DC field applied perpendicular to the $c$-axis,~\cite{Francisco_PRB_17_FMR} with Kittel-like modes with a square-root of frequency dependence~\cite{kittel1948} at high field magnitudes ($\gtrsim 150$~mT) when the sample is in the F-FM phase, and asymmetric CSL-phase resonances at lower fields.
The multiple F-FM resonances are related to the existence of a size dependent inhomogeneous demagnetisation field.~\cite{FTG_PRB18_tailored_FMR}
In the larger sample used here, the CSL modes resemble a rotated sigmoid and the resonances are more complex, with the number of resonance branches varying greatly as the field is swept and the magnetisation transitions between different phases.

The end of a clear single domain F-FM phase with decreasing field occurs around +136~mT, while the transition from the CSL phase to a the F-FM one occurs at -156~mT, as indicated in the figure annotations.
Due to the FMR sample being much larger than the sample used for the dislocation densities measurements shown in Figure~\ref{fig:fresnel_dislocation_populations}, the critical F-FM fields measured in the FMR sample are relatively similar.
However, they are not identical, and the discrepancy of the values is ascribed to the existence of the surface barrier in the sample, as discussed in the literature.~\cite{Mito_PRB_18_geometrical_protection, Shinozaki_PRB18_soliton_barrier}
Only in bulk samples (0.1-1~mm along the $c$-axis)~\cite{togawa_2013_PRL_MR} are discretisation related hysteresis effects absent.

The co-existence of CSL and F-FM regions in the F-FM to CSL transition is readily apparent from the overlapping Kittel-like and CSL-like resonances at positive fields below 136~mT.
The collective resonance properties of each of these regions will be determined by the region size and on the surrounding magnetisation state which, in turn, are defined in part by the dislocation distribution.
The dislocations may act as pinning sites or spin wave scatterers through distortion of the lattice, which at low fields can extend over 100~nm in range (see Figure~\ref{fig:dislo_period}(b)).
Indeed, modelling of collective resonances in a confined CSL shows that the boundary conditions of the lattice and its soliton chain length influence the resonance frequency,~\cite{Kishine_PRB_16_pined_CSL_resonance} with lower pinning fields and shorter chains having lower frequencies.
We may thus tentatively ascribe the multiple low amplitude and low frequency FMR branches observed here to a combination of a distribution of different CSL chain lengths, CSL region sizes and boundary conditions, and F-FM regions, all defined by dislocations; in effect, the sample may be partitioned into smaller magnetic regions, each with a different low amplitude precessional mode.
As the field is reduced, the progression of dislocations through the system modifies the local boundaries of the effective volumes contributing to a given mode. 
In this picture, when one or more dislocations defining a region vanishes, the small amplitude resonances from the associated volumes increase in frequency or disappear altogether while simultaneously increasing the amplitude of the main CSL modes.
The sharp, step-like nature of the frequency change and the high degree of synchronisation (\textit{e.g.} aligned jumps in frequency marked by vertical arrows in Figure~\ref{fig:fmr}) is consistent with the dislocation generation and movement which, as discussed earlier, occurs very rapidly in uniform thickness samples.
Only when a small negative field is reached are all dislocations are finally removed from the system, consistent with the TEM data in Figure~\ref{fig:fresnel_dislocation_populations}, leaving it in the ordered CSL phase with a macroscopic coherence length, and only the main CSL resonance modes are observed.

Because expulsion of solitons from an ordered CSL phase is a coherent process which maintains the long range lattice order, only modes characteristic of a single domain CSL phase are observed until relatively large negative fields are applied and, consequently, the CSL to F-FM transition is much sharper.
At fields just above the CSL to F-FM critical field there occurs a small and gradual redistribution of resonance intensity from the main CSL resonance to lower frequency branches (marked by an `*' in Figure~\ref{fig:fmr}(b)).
While the origin of this behaviour is unclear, it is interesting to note that a small number of dislocations are observed in the equivalent TEM data at this stage of reversal (see Figure~\ref{fig:fresnel_dislocation_populations}) and, thus, these may also influence the formation of CSL and F-FM regions in the CSL to F-FM transition.

Quantisation effects and the macroscale coherence of the CSL phase are key properties of the CrNb$_3$S$_6$ system for potential application in areas of spintronics and nanomagnetism and, so, it is important that further theoretical work is performed to understand in greater detail the role of disorder in these effects and also the influence of surface spins on this interesting and highly configurable magnetic system.

\subsection*{Guided Motion in Dislocation Mediated Phase Transitions}
\begin{figure}[!htb]
  \centering
    \includegraphics[width=4.5cm]{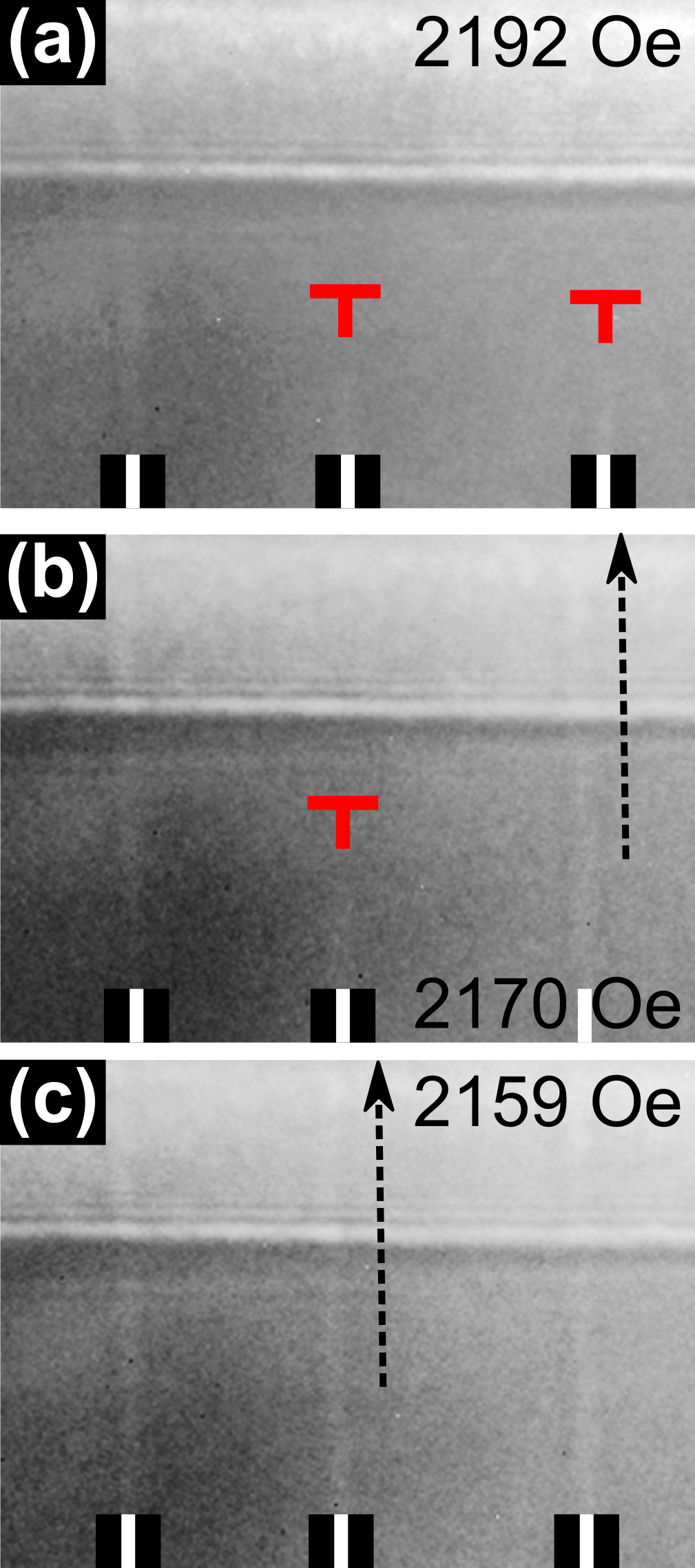}   
    \caption{{Fresnel imaging of solitons running vertically in a thin lamella, showing unidirectional guided motion of soliton dislocations (red $\perp$ symbols symbols) in response to a decreasing (a)-(c) applied magnetic field during the F-FM to CSL phase transition. The two different background intensities are due to thickness variations in the sample.}
    \label{fig:guided_motion}}
\end{figure}

Finally, evidence for field driven dislocation movement modulating CSL growth is presented in the Fresnel images of the magnetic state at different stages of the F-FM to CSL phase transition depicted in Figure~\ref{fig:guided_motion}.
Dislocations, marked by the red $\perp$ symbols, move upwards following the black arrows as the field is decreased in panels (a) to (c), enlarging the CSL area.
In following the lowest energy path, the dislocation movement allows each soliton to follow the surrounding lattice in a `guided' manner, irrespective of crystal thickness.
Interestingly, this guided movement persists even in very dilute CSL phases, as shown in Figure~\ref{fig:guided_motion_sup}.
Just as large crystal thickness steps stabilise dislocations, this field driven guided dislocation motion may be made possible by the thickness modulation in gradually tapered samples such as that in Figure~\ref{fig:guided_motion_sup}, further demonstrating the robust nature of solitons in CrNb$_3$S$_6$ and the ability to tailor their properties through changes in sample morphology.
Control of the dislocation position and the size of the CSL phase in this way may be exploited in device applications such as in guiding microwave fields or in field dependent microwave attenuators.

\section*{Conclusions}
We have experimentally characterised the magnetisation in the CrNb$_3$S$_6$ system at multiple length scales, from nanometer-scale resolved DPC images of chiral soliton lattice distortion at dislocations, through micron-scale Fresnel imaging of dislocation populations and lattice parameter during magnetisation reversals, to the tens of microns scale with FMR measurements of the dynamic magnetisation properties.
The formation and movement of soliton dislocations is found to mediate the formation of F-FM and CSL regions and this can be highly hysteretic.
The dislocations strongly influence the magnetisation reversal and the resulting coherence of the magnetic configuration, knowledge of which is critical to understanding and controlling the magnetisation properties.
Sample morphology is found to play a key role in stabilising normally unstable or metastable dislocations and allows the magnetic properties to be controlled and modified far beyond that predicted for a uniform sample, potentially allowing the use of the tunnable emergent nano-channels of field polarised spins in device applications.

\section*{Methods}
\subsection*{Simulations}
The mean-field simulation in Figure~\ref{fig:numerical_simulations} was performed for a three-dimensional lattice with a spin Hamiltonian which consists of ferromagnetic exchange interactions ($J_\parallel$ for bonds parallel to the $x$-axis, $J_\perp$ for bonds perpendicular to the $x$-axis), DM interaction for bonds parallel to the $x$-axis (coupling constant $D$), easy-plane anisotropy (coupling constant $K_u$) and Zeeman energy due to external field almost parallel to $z$-axis.

In the mean-field approximation, local spin $\bm{S}_i$ (with modulus $S$) on the site $\bm{i} = (i_x, i_y , i_z)$ $\in$ [1, $N_x$] $\otimes $ [1, $N_y$] $\otimes$ [1, $N_z$] is subject to the mean field (molecular field) $\bm{H}_i^{eff}$ and the expectation value $\langle\bm{S}_i\rangle$ is given by $\langle\bm{S}_i\rangle = S f(S|\bm{H}_i^{eff}|/(k_{B}T)) \bm{H}_i^{eff}/|\bm{H}_i^{eff}|$ with $f(x)=\coth(x-1/x)$.
$\bm{H}_i^{eff}$ is given as a function of $\{\langle\bm{S}_j\rangle\}_j$ and thus the mean field theory yields a self-consistent equation for  $\{\langle\bm{S}_j\rangle\}_j$.

The iteration procedure to seek for solution to this self-consistent equation consists of the following steps: (i) prepare an initial state $\{\langle\bm{S}_j\rangle\}_j$;  (ii) calculate $\bm{H}_i^{eff}$ for a given $\{\langle\bm{S}_j\rangle\}_j$;  (iii) update $\langle\bm{S}_i\rangle$ by $S f(S |\bm{H}_i^{eff}|/(k_{B}T)) \bm{H}_i^{eff}/|\bm{H}_i^{eff}|$; (iv) repeat (ii) and (iii) until $\{\langle\bm{S}_j\rangle\}_j$ converges.
The updated series of $\langle\bm{S}_i\rangle$ can be regarded as a kind of relaxation process.

For the present purpose, we focus on a transient time domain, when the dislocations exhibit slow dynamics to annihilation.
We take as an initial state with $\bm{S}_i = S(\cos(2\pi w i_z/N_z),  \sin(2\pi w i_z/N_z), 0)$ for $i_y \in [1, N_m]$ and $\bm{S}_i = S(\cos(2\pi(w+1) i_z /N_z), \sin(2\pi(w+1) i_z/N_z), 0)$ for $i_y \in [N_m+1, N_y]$ with an integer $N_m \in (1, N_y)$.
We take $(N_x, N_y, N_z) = (200, 200, 5)$.
The integer $w$ denotes a winding number.
Typically, it takes 10000 iterations to relax a state with a pair of dislocation and 80000 iterations to annihilate the pair.

When presented the projected components of the magnetisation, the spacing between spins was taken as 0.575~nm in $x$, and 1.21~nm in $y$.~\cite{Miyadai_JPSJ_1983}

\subsection*{STEM Sample Preparation and Measurements}
The STEM sample was prepared from a CrNb$_3$S$_6$ crystal in an FEI Nova NanoLab DualBeam focused ion beam (FIB), and thinned to electron transparency using standard lamella preparation procedures.

Differential phase contrast (DPC) and electron energy loss spectroscopy (EELS) measurements were performed in a probe corrected JEOL ARM 200F scanning transmission electron microscope equipped with a cold field emission electron gun operated at 200~keV.
In the DPC measurements, the sample was cooled to $\sim$102~K in a Gatan HC3500 sample holder while the signal was collected by a custom 8-segment detector,~\cite{mcvitie2015_magtem} allowing the modified DPC method to be used to reduce diffraction artefacts.~\cite{chapman_1990_mdpc}
During the measurements, the field was applied in a direction perpendicular to the sample and varied by exciting the objective lens to different strengths.
The convergence semi-angle was 1.31~mrad, giving a spot size of 2.34~nm and a resolution of 1.17~nm; the pixel size was 1.24~nm for the low field data and 0.74~nm for the high field data.

The sample thickness was determined from a $t/\lambda$ map~\cite{Malis1988_tol} obtained from EELS measurements performed in the same microscope with the objective lens switched on, and an effective Z-number of 25.8, calculated using Gatan Digital Micrograph.
The EELS spectra were collected using a GIF Gatan Quantum ER spectrometer in scanning TEM (STEM) mode.
The convergence and collection semi-angles were 29 and 36~mrad, respectively.

In Figure~\ref{fig:dpc_dislocation}, the dimension in (e) is the average over the three left-most solitons shown in (d), while that in (f) is the average of the two profiles along the maxima (red) and minima (blue) of the high field dislocation.
The errors quoted are from the fits to the data with an additional error estimated from a calibration measurement, except in (e), where the value and error are the mean and standard deviation of the measurements of the three solitons shown.

All data were processed using Python and the FPD~\cite{fpd} and HyperSpy~\cite{hyperspy} packages.

\subsection*{TEM Measurements}
In order to see the CSL behavior with dislocations qualitatively, Fresnel images were recorded in underfocus condition (typically $\sim$1~$\muup$m defocus) using a model JEM-2010 or JEM-2100 transmission electron microscope (TEM) operated with an acceleration voltage of 200~kV.
The TEM samples were cooled using a liquid N2 holder for the experiments performed in a temperature range above 100~K.
To see the behavior at lower temperatures for many hours, the seventh generation cryogenic transmission electron microscope, available in Nagoya University, Japan~\cite{Yoshinori_2011_cryo_tem} was adopted.
The samples were cooled to below 2~K using superfluid He and the CSL behavior was examined at low temperatures in magnetic fields systematically, as described in the main text.

\subsection*{FMR Measurements}
FMR data was recorded using a vector network analyser (VNA) coupled to a broadband coplaner waveguide with the sample cooled to 50~K.
The transmitted power ($S_{21}$) through the ground-signal-ground waveguide to which the sample was attached (see Figure~\ref{fig:fmr_sup}) was recorded as a function of frequency as the applied field was reduced from a large positive value ($|H|>H_c$) to a similarly large negative value.

The attenuation of the transmitted signal is strongly frequency dependent and so we normalise the data at each frequency to a high percentile value of the field dependent data across all fields.
This procedure avoids resonances in a reference spectra appearing in the normalised data and, minimising the influence of this type of artefact.
The data is further filtered by applying median line correction to correct small variations in scattering parameter at different frequencies due to small drifts in the VNA electronics.

\section*{Acknowledgements}
This work was supported by the Engineering and Physical Sciences Research Council (EPSRC) of the UK under grant EP/M024423/1 and the JSPS Core-to-Core Program `Advanced Research Networks'.
We acknowledge support from Grants-in-Aid for Scientific Research (No. 25220803, No. 17H02767, No. 17H02923) and JSPS Core-to-Core Program, A. Advanced Research Networks. 
This work was also supported by Chirality Research Center (Crescent) in Hiroshima University.
F.G. received additional support from the JSPS International Research Fellowship No. 17F17316.
G.W.P. thanks William Smith for preparing the STEM lamella.
Y. T. and T. K thank Y. Fujiyoshi for the opportunity to use the seventh generation cryogenic transmission electron microscope suitable for TEM observations at low temperatures for many hours.
Original data files are available at DOI TBA.

\section*{Author Contributions}
G.W.P. calculated the simulated DPC signals, performed the STEM experiments with assistance from M.N., analysed the data, and wrote the manuscript with contributions from F.G., Y. Kato, S.McV., and Y.T.;
T.K. performed the TEM measurements and analysis under the supervision of Y.T.;
M.S. performed the mean-field calculations with theoretical and numerical contributions from Y.M. and under the supervision of Y. Kato;
F.G., T.S. and Y.S. performed the FMR measurements;
T.S. and Y.S. fabricated the FMR sample;
Y. Kousaka performed the crystal growth and quality examination;
All authors reviewed and commented on the manuscript.

\section*{Additional information}
\textbf{Competing Interests}

\noindent The authors declare no competing financial interests.

\section*{Supplemental Information}
Supplementary materials are available at DOI TBA.

Examples of simulated DPC images, measured dislocation induction components, sample overview images, dislocation induced lattice distortion, reversal properties in a confined geometry, examples of dislocation instability, details of the FMR setup, and evidence of dislocation guided motion in dilute CSL phases (pdf).


\onecolumn
\renewcommand{\thefigure}{S\arabic{figure}}
\setcounter{figure}{0}

\cleardoublepage
\part*{Supplemental Information for `Order and Disorder in the Magnetisation of the Chiral Crystal CrNb$_3$S$_6$'}

\section*{Simulated DPC Images}
\label{sec:simulated_dpc_sup}
As discussed briefly in the manuscript, divergent components of the magnetisation perpendicular to the electron trajectory are not imaged by the differential phase contrast (DPC) technique.
In fact, it is magnetisation components with a net curl that are imaged.~\cite{Stephen_JAP_2001} 
To understand the differences in the magnetisation and the induction produced by it, it is therefore instructive to examine the components contributing to the divergence and curl of the simulated magnetisation distribution, $M$, discussed in the manuscript and plotted in Figure~\ref{fig:numerical_simulations}. 

\begin{figure*}[!htb]
  \centering
    \includegraphics[width=17.5cm]{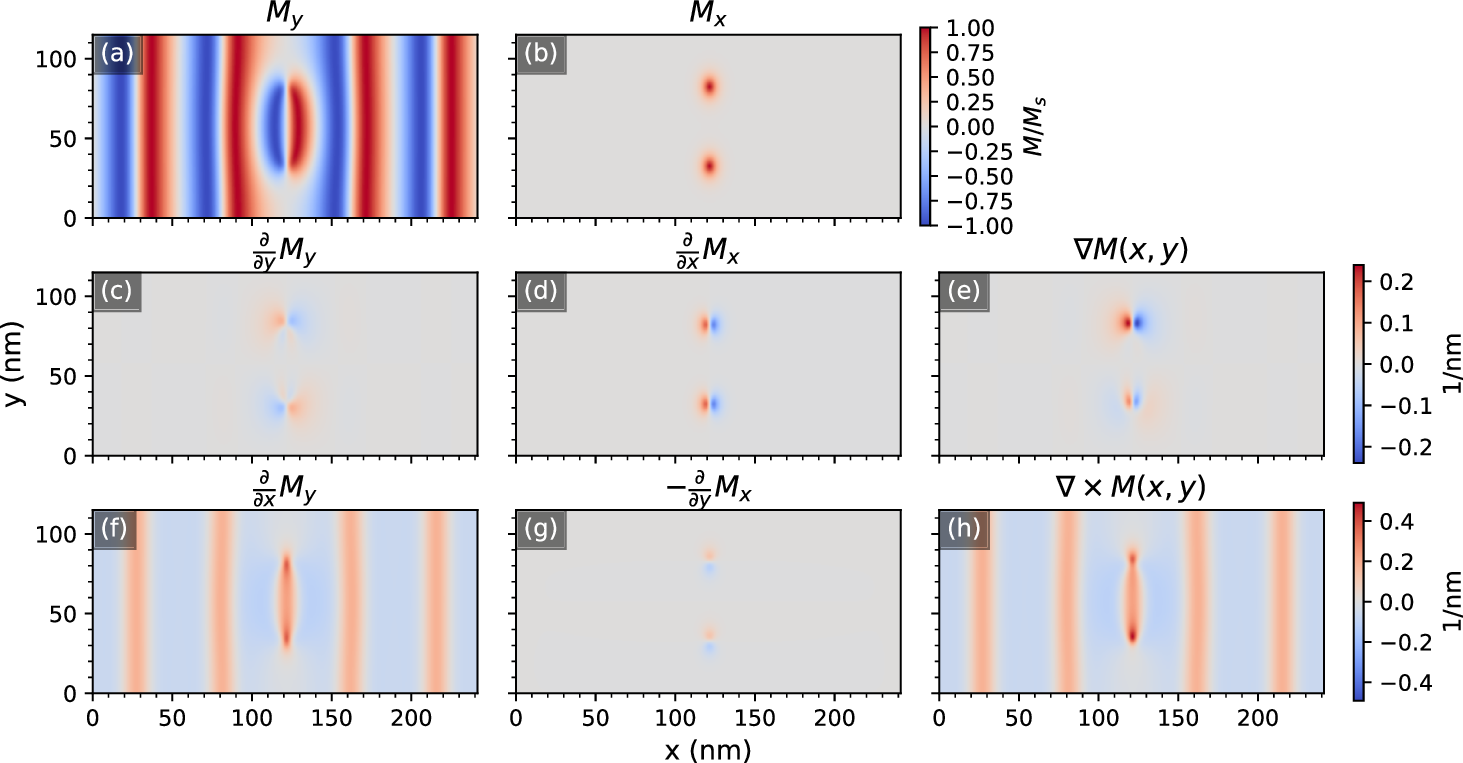}   
    \caption{{Normalised in-plane simulated magnetisation components (a, b), and the calculated components and total value of the divergence (c-e) and curl (f-h) of the magnetisation.}
    \label{fig:simulation_m_div_curl}}
\end{figure*}

Figures~\ref{fig:simulation_m_div_curl}(a) and (b) show the normalised simulated magnetisation components, with blue being negative, grey zero, and red positive, in the directions indicated by the axes.
In the second row are the divergence components (c-d) and the total divergence (e), as indicated by the annotations.
Similarly, the components (f-g) and total (h) of the magnetisation curl are shown in the third row.

What is clear from the divergence images is that both the $x$- and $y$-component give a divergence (charge density) that is dipolar in character, with a moment aligned parallel to the $x$-axis.
The dipole moments from the $x$-component are aligned parallel to one another (d), while those from the $y$-component are anti-parallel to one another (c), and the moments are much stronger from the $x$-component compared to the $y$-component.
This means that the total divergence (e) is additive at the top of the dislocation but subtracts at the bottom one, with the result that the moment for each resultant dipole is in the same direction, though of a different magnitude. 
Therefore, the difference between magnetisation and induction would be expected to be smaller at the lower dislocation compared to the top one.

The curl of the $x$-component of the magnetisation (Figure~\ref{fig:simulation_m_div_curl}(g)), has a small net value at the locations of the dislocation, and these are also of the opposite sign as the much stronger curl of the $y$-component of the magnetisation (f).
As a consequence, the integrated induction maps well the $y$-component of the magnetisation, with small and asymmetric differences at the locations of the dislocations, as discussed next.

\begin{figure*}[!htb]
    \includegraphics[width=16.0cm]{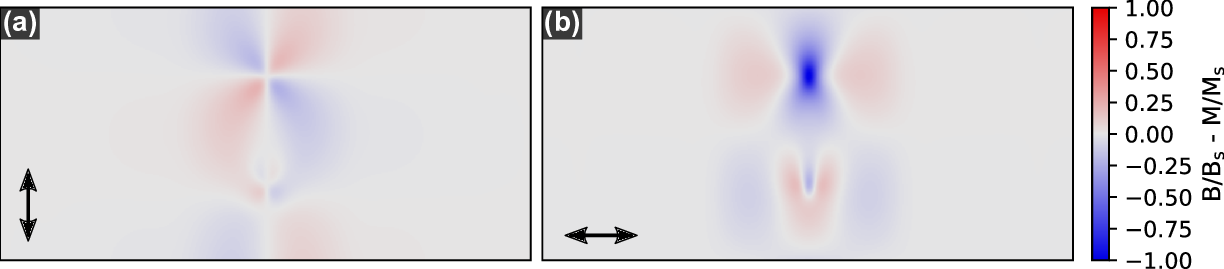}  
    \caption{{Difference in the components of projected magnetic induction imaged by DPC and the magnetisation from simulated dislocations. The arrows indicate the magnetisation component orientation.}
    \label{fig:numerical_simulation_HM_error}}
\end{figure*}

\begin{figure}[!htb]
  \centering
    \includegraphics[width=10.0cm]{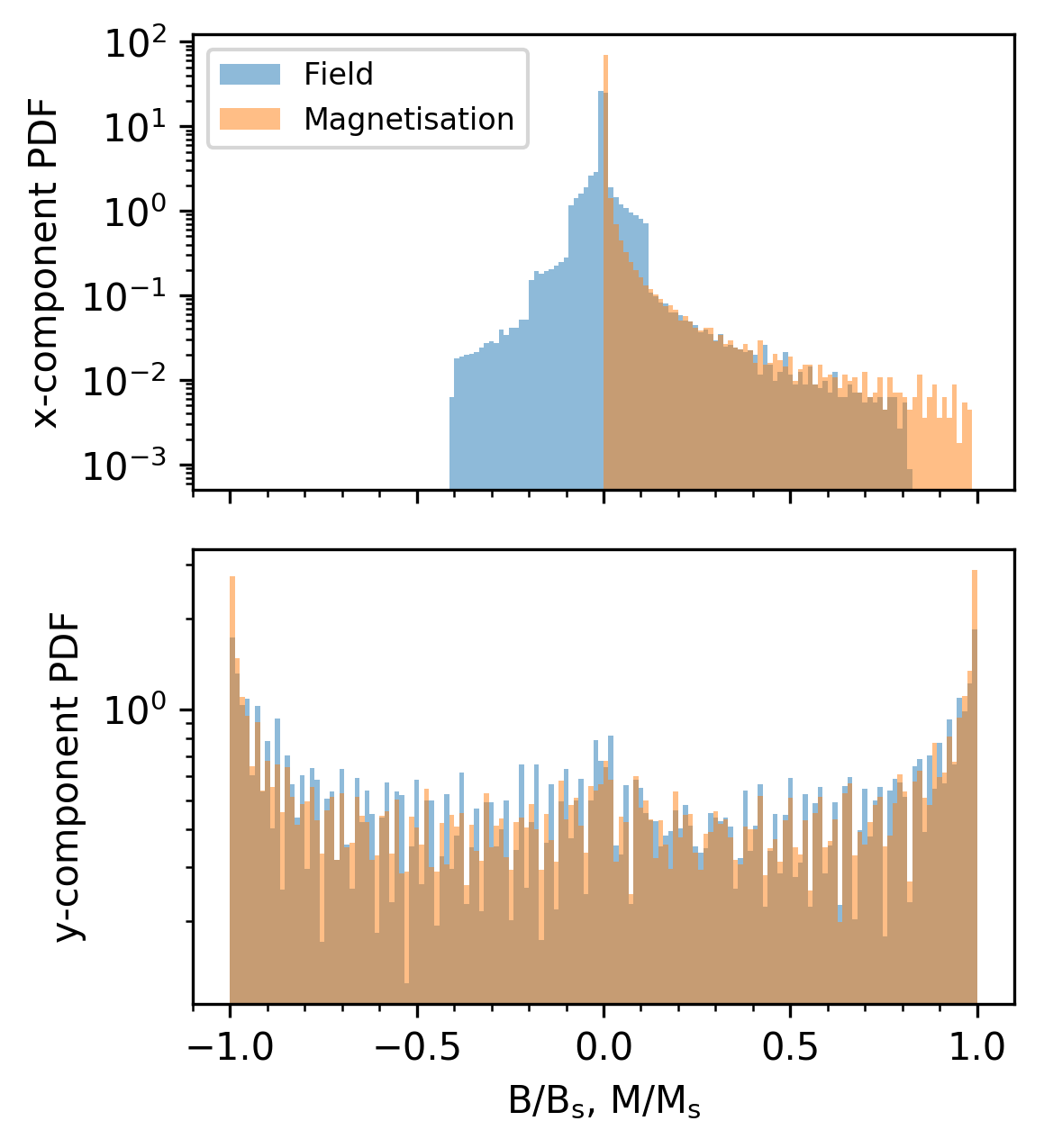}   
    \caption{{Normalised probability distributions for the projected components of the simulated dislocation magnetisation and of the induction field distributions imaged by DPC, in normalised units.}
    \label{fig:numerical_simulation_histograms}}
\end{figure}

Figures~\ref{fig:numerical_simulation_HM_error} and \ref{fig:numerical_simulation_histograms} compare the integrated magnetic inductions, $B$, predicted to be produced from DPC imaging of the simulated magnetisation distribution discussed in the manuscript and above.

Figure~\ref{fig:numerical_simulation_HM_error}(a) and (b) show the differences in the normalised magnetisation and inductions for the $y$- and $x$-components, as indicated by the arrows, calculated from the data in Figure~\ref{fig:numerical_simulations} (b) and (d), and (c) and (e), respectively.
It should be pointed out that this, therefore, represents the integrated magnetic field intensity, $H$, which is added to the projected magnetisation to obtain the DPC images when converted to induction ($B = \mu_o H$).
This contrast is consistent with two dipoles located at either end of the dislocation pointing in the same direction but with the upper dipole having a larger moment and therefore field, as discussed above.
We note that there is no preferred direction to the $x$-component of magnetisation from the dislocation and the weaker dipole will have the lowest magnetostatic energy and therefore might be the most likely one to be observed.
In any case, these images explain the contrast predicted in Figures~\ref{fig:numerical_simulations}(d) and (e) of the main text, where the $y$-component is dominated by the magnetisation, whereas the resultant induction for the $x$-component is overall weaker.
It may be expected that the induction component parallel to the $c$-axis ($x$-axis in these figures) may be challenging to image by DPC at dislocations.

Figure~\ref{fig:numerical_simulation_histograms} shows a histogram of the $x$- (top) and $y$-component (bottom) of $M$ and $B$.
The range and distribution of $y$-components of $M$ are well measured by DPC, but the $x$-components are not, with the range, sign and magnitude distributions changing significantly due to the divergence of the dislocation magnetisation.

\FloatBarrier
\section*{Measured Dislocation Induction Components}
\begin{figure*}[!htb]
  \centering
    \includegraphics[width=12.5cm]{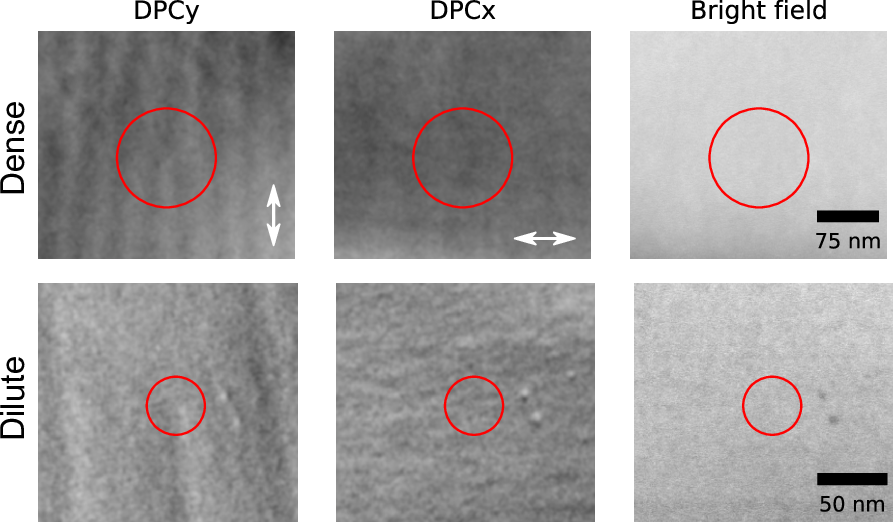}   
    \caption{{The raw $y$- and $x$-components DPC signal for the dense and dilute CSL images with dislocations (circled) shown in Figure~\ref{fig:dpc_dislocation}(a) and (d), respectively. The arrows indicate the direction of the components measured. The components of each DPC dataset are plotted on the same deflection scale. The rightmost column shows the bright field images from the same datasets.}
    \label{fig:induction_components}}
\end{figure*}

Figure~\ref{fig:induction_components} shows the raw $y$- and $x$-components of the DPC signal (without thickness normalisation) for the dense and dilute CSL images with dislocations shown in Figure~\ref{fig:dpc_dislocation}(a) and (d), respectively, along with the bright field images produced from the same dataset.
Background variations in the DPC data in the manuscript from FIB curtaining and contamination were minimised through subtraction of field polarised data and further processing to reduce the influence of bend contours.
The DPC data in Figure~\ref{fig:induction_components} only has a plane subtracted and a 1.5 pixel wide Gaussian blur applied to reduce the high frequency components of noise.
The positions of the dislocation in each image is marked by red circles.
The dark spots in the bright field images for the high field (dilute) data are from carbon build up and do not affect the magnetisation.

We note that in both datasets, no $x$-component is observable within the achieved signal-to-noise level.
This may be in part because of the reduced sensitivity of the imaging technique to this particular component, as discussed in relation to Figure~\ref{fig:numerical_simulation_HM_error} and, potentially, because the dislocation may be dynamic, as discussed in relation to Figure~\ref{fig:sup_dpc_slow_dynamics}.

\FloatBarrier
\section*{Sample Overview}
\begin{figure*}[!hbp]
  \centering
    \includegraphics[width=7.5cm]{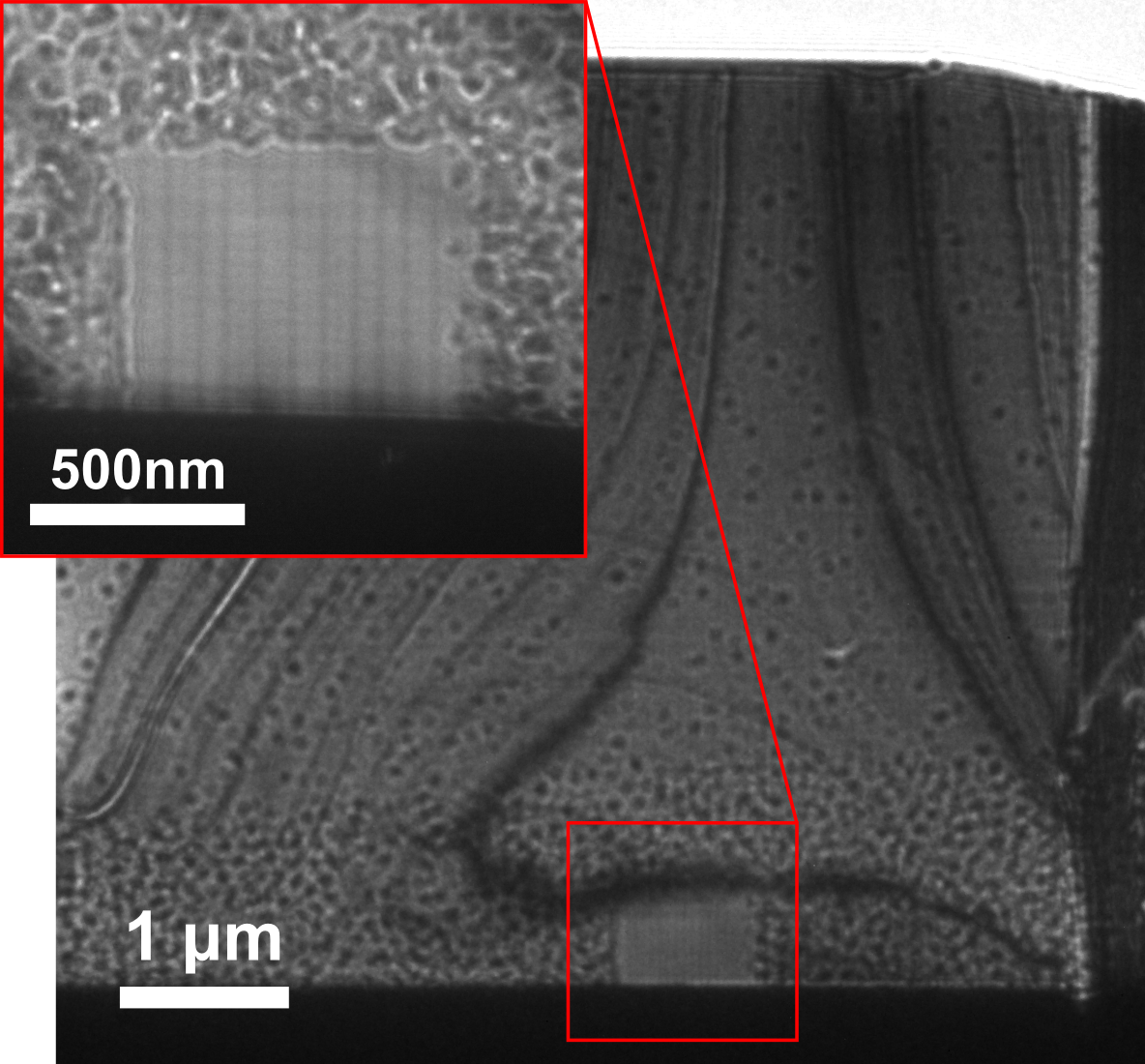}   
    \caption{{Pseudo Fresnel images showing an overview of the sample at 102~K, produced in STEM mode by defocusing the probe. The inset is an enlarged view of the region highlighted by the red box, imaged under a slightly different sample tilt. The approximate area imaged in the experiments is indicated by the contamination-free area.}
    \label{fig:sample_overview}}
\end{figure*}

Figure~\ref{fig:sample_overview} shows an overview of the sample used in the study, imaged at a temperature of 102~K.
At this temperature, low levels of contamination in the vacuum of the microscope adsorb onto the sample and this can be seen in the `mottled' contrast of the otherwise uniform film with bend contours. 
The approximate area imaged in the DPC studies is indicated by the contamination-free area, and is located adjacent to the large increase in sample thickness, as discussed in the main text.   

\FloatBarrier
\section*{Dislocation Induced Soliton Lattice Distortion}
\begin{figure*}[!htb]
  \centering
    \includegraphics[width=13cm]{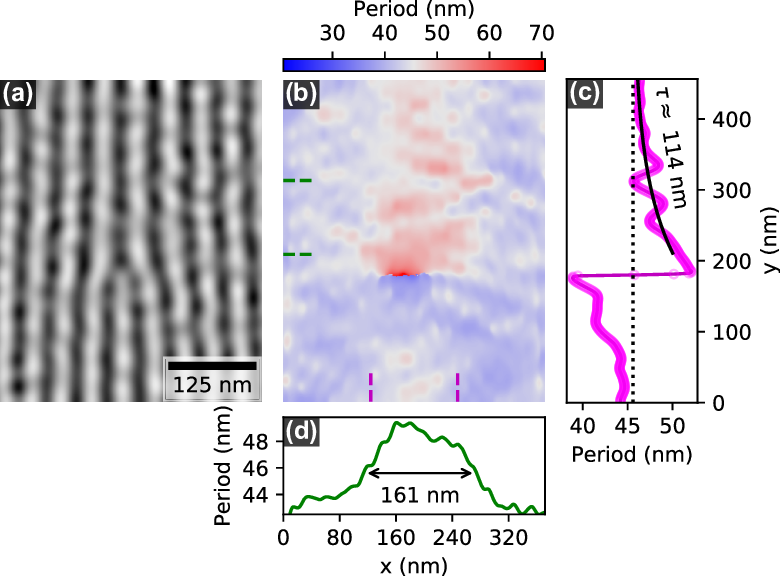}
    \caption{{Analysis of the CSL deformation around a low field dislocation in the DPC data in Figure~\ref{fig:dpc_dislocation}(a). (a) Fourier filtered data. (b) Local CSL periodicity obtained by fitting a sinusoidal function to overlapping 88~nm sections along each row of the filtered data. Note that the fitting width limits the accuracy in close proximity to the dislocation, but otherwise accurately measures the magnitude and extent of the lattice distortion. (c) and (d) show periodicity profiles from the average of the areas delimited by the similarly coloured dashed lines in (b). The black line in (c) is an exponential fit to the decay in period perpendicular to the $c$-axis, while the dimension in (d) is the approximate full width half max value of the distortion parallel to the $c$-axis.}
    \label{fig:dislo_period}}
\end{figure*}

Figure~\ref{fig:dpc_dislocation}(a) of the main article shows the induction produced by the $y$-component of magnetisation of a sample with a dislocation, measured at low fields with DPC.
As shown in Figure~\ref{fig:dpc_dislocation}(b), the periodicity of the chiral soliton lattice (CSL) varies in the region of the dislocation.
To map the distortion in two dimensions, we Fourier filtered the same data and then fitted sinusoidal functions to overlapping regions across the entire scan.
The results of this are shown in Figure~\ref{fig:dislo_period}(a) and (b), where the colour map in the period data in (b) is centred around 45.6~nm (grey), with blue indicating shorter periods, and red longer ones.
The magnetic lattice is most strongly modified at the location of the dislocation, as one would expect, but distortion of the lattice extends for 10s to 100s of nanometres, demonstrating the significant disturbance to the surrounding lattice created by a single dislocation.

The period profiles from the average of the regions marked by the dashed lines in Figure~\ref{fig:dislo_period}(b) are shown in (c) and (d) for the directions perpendicular (magenta) and parallel (green) to the $c$-axis, respectively.
The full width half maximum value of the distortion parallel to the $c$-axis is around 160~nm.
To estimate the distortion lengthscale in the perpendicular direction, we fitted a decaying exponential function (black line) to the data (magenta) in (c) and extracted a decay constant of around 114~nm.

The mean period value used in the above analysis was extracted from the average period profile perpendicular to the $c$-axis shown in Figure~\ref{fig:dislo_period}(c) by fitting the simple function $a/(y-y_o) + p_o$ to the data, where $a$ and $y_o$ are constants and $p_o$ is the predicted period far from the dislocation.
Such $1/y$ expressions arise in models of stress in 2-D crystals with edge dislocations,\cite{Physical_Metallurgy_Principles} but we use it in our 1-D magnetic lattice only to estimate the value of $p_o$ in the measurement.
This mean value is also indicated by the dashed black vertical line in (c).
In the fit, we excluded the area within 30~nm of the dislocation to avoid inaccuracies arising from using the average profile.

\FloatBarrier
\section*{Reversal in a Confined Geometry}
\begin{figure*}[!htb]
  \centering
    \includegraphics[width=12cm]{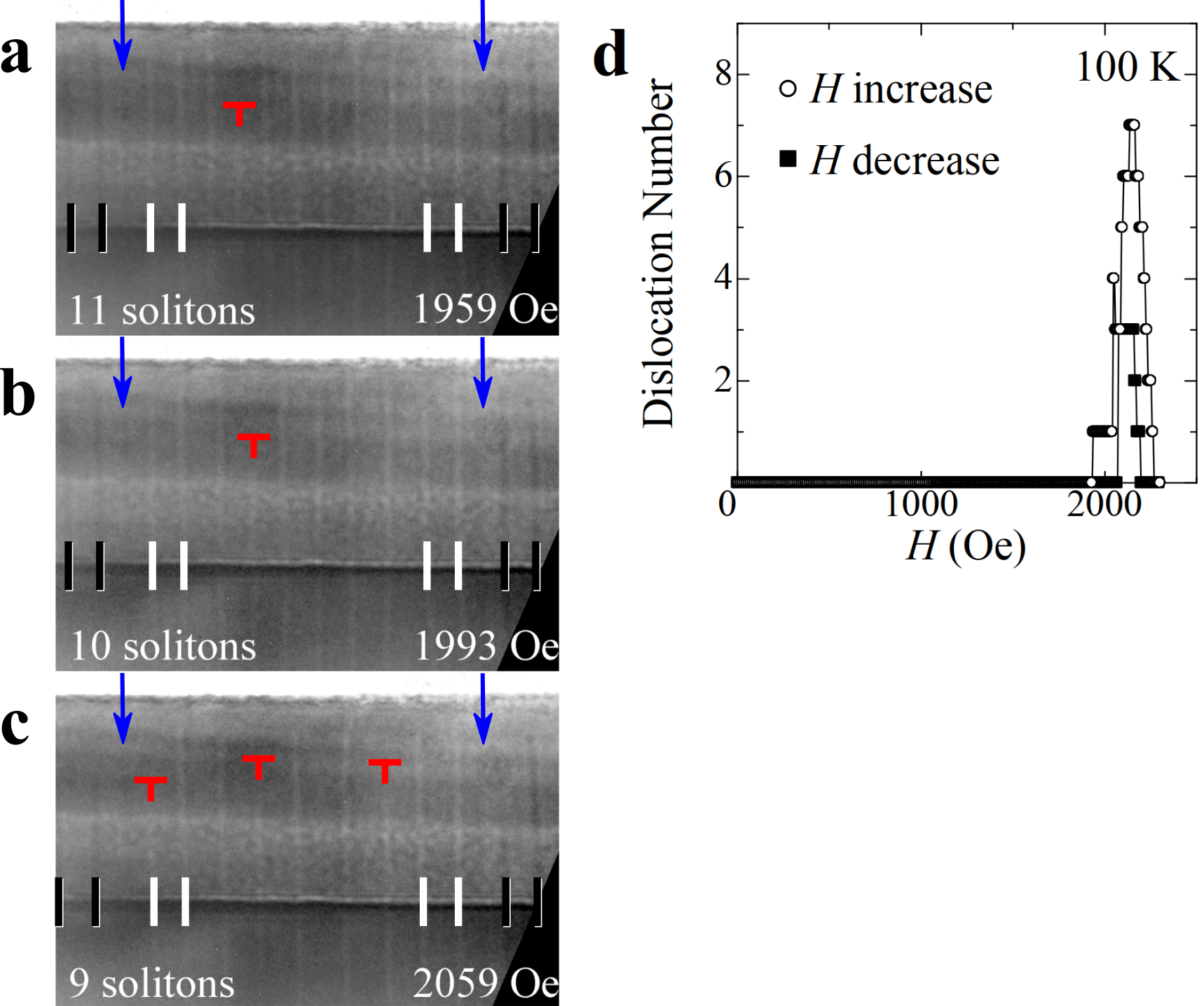}
    \caption{{(a) - (c) Nucleation of dislocations, marked by red $\perp$ symbols, in a confined right-hand chirality grain in a left-handed crystal, in response to an increasing applied field strength. The grain boundaries are indicated by the blue arrows. The variation in background intensity is due to sample thickness variations. (d) Dislocation number observed in both sweep directions.}
    \label{fig:nucleation_in_grain}}
\end{figure*}

The central sections of the Fresnel images in Figure~\ref{fig:nucleation_in_grain}, bounded by the blue arrows, is a left-handed grain of opposite chirality to that of the adjacent right-handed material, and the soliton contrast is inverted in the two regions because of this.
The edge of the sample is visible at the top of each panel, while the darker section in the bottom third of each panel is due to increased sample thickness.
Dislocations are marked by red $\perp$ symbols and appear to nucleate at the edge of the sample and move downward with increasing field, towards the thicker sample region, as shown in Figure~\ref{fig:nucleation_in_grain}.

The number of solitons within the grain decreases through panels (a) to (c) as the field is increased and solitons and the dislocations in them redistribute by moving laterally to accommodate the lower density.
This can be easily seen in the single dislocation in panel (a) moving to the right, along the $c$-axis, in panel (b) when a single soliton is removed.
More dislocations form as the field is further increased in panel (c).
Panel (d) shows the dislocation density as the field is swept in both increasing and decreasing directions and, while the number of dislocations is rather low, a similar number of them are observed in both datasets and they both peak in number at high field.

The lack of dislocation hysteresis in this confined volume may be due to reduced number of possible nucleation sites and the step increase in sample thickness, as discussed in the manuscript.
Additionally, the presence of grain boundaries may play a role.
At grain boundaries, the magnetisation is pinned in-plane and perpendicular to the $c$-axis in opposite directions at each interface.~\cite{Yoshi_PRB_15_confinement}
Because of this, the angle through which moments adjacent to the boundary must rotate to generate a soliton compared to those in the F-FM regions is reduced, and may decrease the energy barrier to soliton generation.
Indeed, crystal lattice grain boundaries have been shown to influence topological magnetic configurations in other chiral systems.~\cite{li_2017_FeGe_dislocation_crystal}
Further theoretical work is required to better understand the role of the interfaces in the CrNb$_3$S$_6$ system.

\FloatBarrier
\section*{Dislocation Instability}
\label{sec:dynamic_sup}
\begin{figure*}[!htb]
  \centering
    \includegraphics[width=10cm]{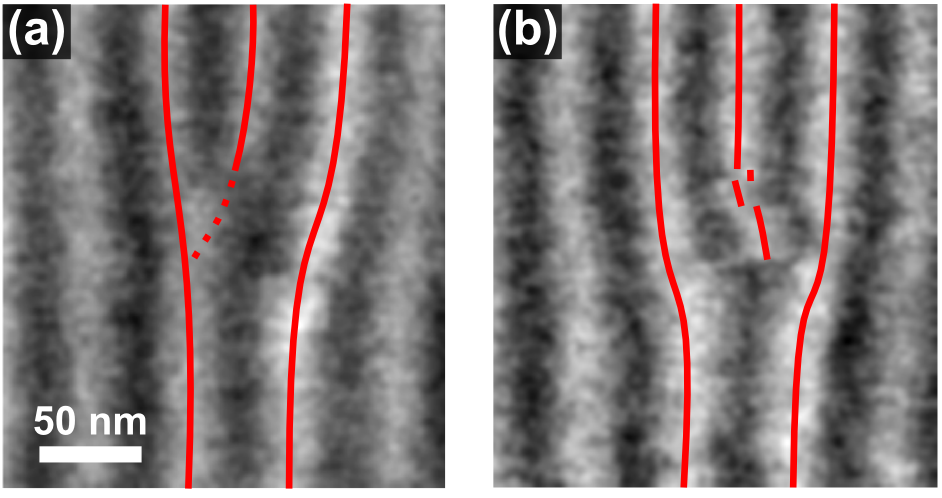}
    \caption{{DPC data taken from a single location showing slow dislocation dynamics at a constant remnant field of 104~Oe and a temperature of 110~K. The soliton dislocation, highlighted with red lines, switches between branches between scans (a-b) and during a single scan (b). The fast scan direction is horizontal and each line takes 0.41~s to complete. The image is 256 by 256 pixels.}
    \label{fig:sup_dpc_slow_dynamics}}
\end{figure*}

The DPC images shown in Figure~\ref{fig:dpc_dislocation} of the main article were from magnetisation configurations that appeared static at a given applied field strength.
As discussed in the manuscript, dislocations are metastable and can exhibit slow dynamics, on the order of the measurement time of a few tenths of a second.
Figure~\ref{fig:sup_dpc_slow_dynamics} shows examples of the slow dynamics of soliton dislocations at low applied field strengths where the CSL is close to the helical state.
The red lines in the DPC scans mark the position of the same dislocation in two sequential scans at the same location.
Between scans, the dislocation switches from curving to the left (a) to curving to the right (b) and, in the second scan, the dislocation can be seen to jump between these two positions during the scan, each line of which took 0.41~s to complete.

We note that while the DPC data in the manuscript appeared static, we cannot rule out faster dynamics than that shown in Figure~\ref{fig:sup_dpc_slow_dynamics} occurring at the dislocation, which could modify the apparent dislocation profile, potentially reducing the DPC signal strength.

\FloatBarrier
\section*{FMR Setup}
\begin{figure}[!htb]
  \centering
    \includegraphics[width=5.5cm]{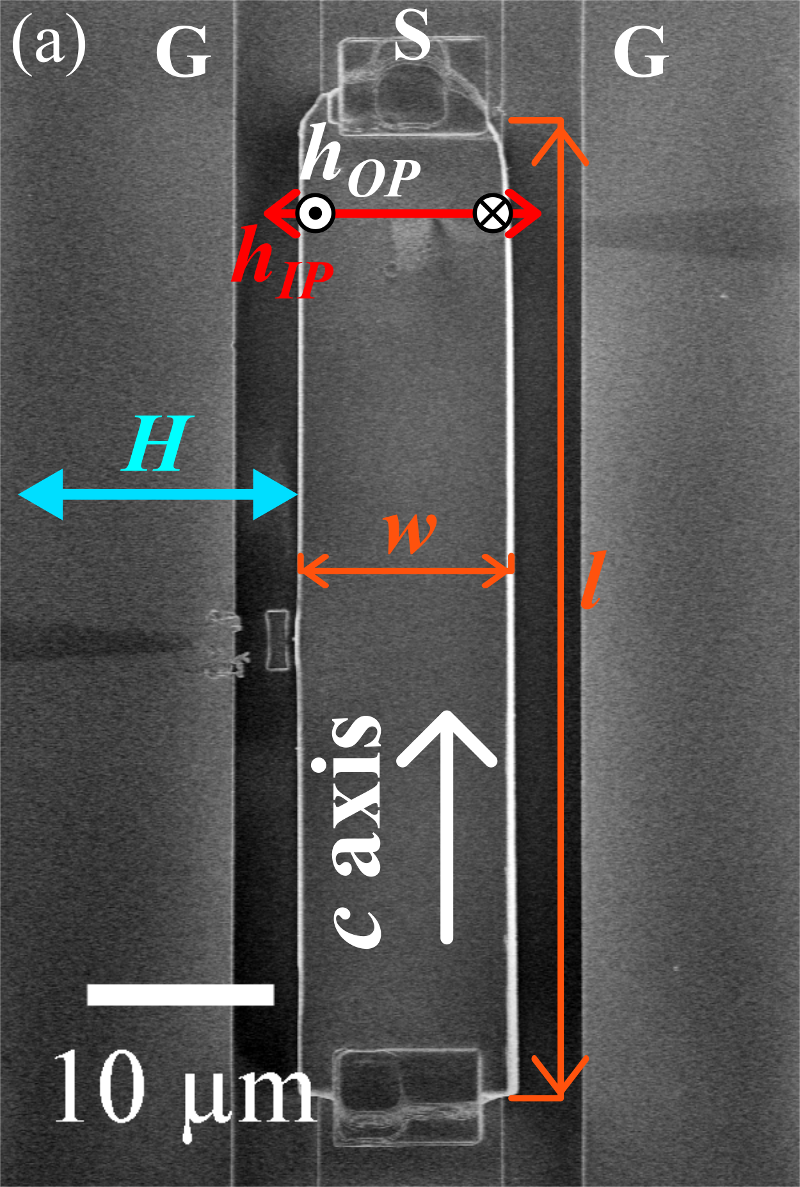}
    \caption{{SEM image of the ferromagnetic resonance experimental configuration, showing the sample mounted on the signal line of the ground-signal-ground (G-S-G) waveguide, the static magnetic field, $H$, and the in- and out-of-plane components of the RF field, $h_{IP}$ and $h_{OP}$, respectively. The sample size was 12.4~$\pm$~0.5~$\muup$m wide, 56.8~$\muup$m long, and 2.6~$\muup$m thick, as defined by the figure annotations.}
    \label{fig:fmr_sup}}
\end{figure}

Figure~\ref{fig:fmr_sup} shows a scanning electron microscope (SEM) image of the ferromagnetic resonance (FMR) sample attached to a coplaner waveguide by electron beam induced platinum deposition, with the $c$-axis aligned parallel to the signal (S) and ground (G) lines.
The DC magnetic field, $H$, and the in-plane, $h_{IP}$, and out-of-plane, $h_{OP}$, components of the radio frequency (RF) field were all applied perpendicular to the $c$-axis of the sample, as shown in the figure. 

The FMR measurements were performed in a custom system described in detail elsewhere~\cite{Francisco_PRB_17_FMR} comprised by a vector network analyser coupled to the broadband waveguide, with the $S_{21}$ scattering parameter measured as a function of frequency at each field.

\FloatBarrier
\section*{Dislocation Guided Motion}
\begin{figure*}[!htb]
  \centering
    \includegraphics[width=11cm]{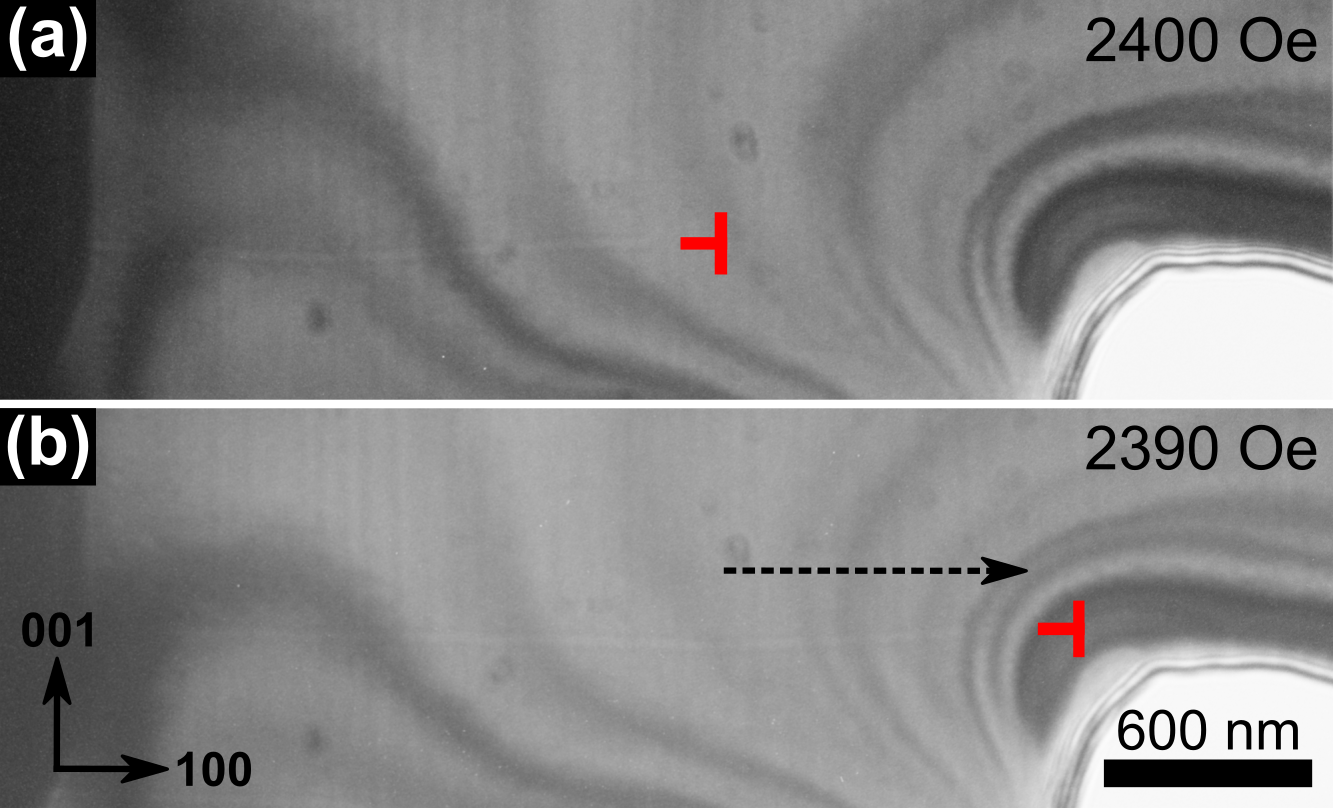}   
    \caption{{Unidirectional guided motion of a soliton dislocation, red $\perp$ symbol, under applied field in a dilute CSL phase imaged in Fresnel mode. The background contrast variation is due to bend contours and thickness contrast as the lamella becomes increasingly thin from the left to the right of the image.}
    \label{fig:guided_motion_sup}}
\end{figure*}

The Fresnel images in Figure~\ref{fig:guided_motion_sup} show an example of the guided motion of a dislocation in a very dilute CSL phase where only one soliton exists in the imaged area.
The dislocation, marked by the red $\perp$ symbol, enters from the left and moves to the right as the applied field is reduced, following an approximately linear path perpendicular to the $c$-axis, showing the robust nature of solitons in CrNb$_3$S$_6$.
This particular feature may be related to the sample thickness, which decreases from left to right, along the dislocation movement direction, creating a shallow wedge shaped sample.
The ability to tailor the normally highly ordered soliton and dislocation properties by changes in sample morphology would potentially allow great customisation of the magnetic properties than is possible in uniform thickness films.

\FloatBarrier

\begin{mcitethebibliography}{36}
\providecommand*\natexlab[1]{#1}
\providecommand*\mciteSetBstSublistMode[1]{}
\providecommand*\mciteSetBstMaxWidthForm[2]{}
\providecommand*\mciteBstWouldAddEndPuncttrue
  {\def\EndOfBibitem{\unskip.}}
\providecommand*\mciteBstWouldAddEndPunctfalse
  {\let\EndOfBibitem\relax}
\providecommand*\mciteSetBstMidEndSepPunct[3]{}
\providecommand*\mciteSetBstSublistLabelBeginEnd[3]{}
\providecommand*\EndOfBibitem{}
\mciteSetBstSublistMode{f}
\mciteSetBstMaxWidthForm{subitem}{(\alph{mcitesubitemcount})}
\mciteSetBstSublistLabelBeginEnd
  {\mcitemaxwidthsubitemform\space}
  {\relax}
  {\relax}

\bibitem[Dzyaloshinskii(1964)]{Dzyaloshinskii_JETP_helicoidal_1964}
Dzyaloshinskii,~I.~E. Theory of Helicoidal Structures in Antiferromagnets. I.
  Nonmetals. \emph{J. Exp. Theor. Phys.} \textbf{1964}, \emph{19}, 960\relax
\mciteBstWouldAddEndPuncttrue
\mciteSetBstMidEndSepPunct{\mcitedefaultmidpunct}
{\mcitedefaultendpunct}{\mcitedefaultseppunct}\relax
\EndOfBibitem
\bibitem[Togawa \latin{et~al.}(2012)Togawa, Koyama, Takayanagi, Mori, Kousaka,
  Akimitsu, Nishihara, Inoue, Ovchinnikov, and Kishine]{togawa2012_fresnel_csl}
Togawa,~Y.; Koyama,~T.; Takayanagi,~K.; Mori,~S.; Kousaka,~Y.; Akimitsu,~J.;
  Nishihara,~S.; Inoue,~K.; Ovchinnikov,~A.~S.; Kishine,~J. Chiral Magnetic
  Soliton Lattice on a Chiral Helimagnet. \emph{Phys. Rev. Lett.}
  \textbf{2012}, \emph{108}, 107202\relax
\mciteBstWouldAddEndPuncttrue
\mciteSetBstMidEndSepPunct{\mcitedefaultmidpunct}
{\mcitedefaultendpunct}{\mcitedefaultseppunct}\relax
\EndOfBibitem
\bibitem[Kishine and Ovchinnikov(2015)Kishine, and
  Ovchinnikov]{KISHINE2015_review}
Kishine,~J.; Ovchinnikov,~A. In \emph{Chapter One - Theory of Monoaxial Chiral
  Helimagnet}; Camley,~R.~E., Stamps,~R.~L., Eds.; Solid State Phys.; Academic
  Press, 2015; Vol.~66; pp 1 -- 130\relax
\mciteBstWouldAddEndPuncttrue
\mciteSetBstMidEndSepPunct{\mcitedefaultmidpunct}
{\mcitedefaultendpunct}{\mcitedefaultseppunct}\relax
\EndOfBibitem
\bibitem[Togawa \latin{et~al.}(2016)Togawa, Kousaka, Inoue, and
  Kishine]{Togawa_jpsj_16_review}
Togawa,~Y.; Kousaka,~Y.; Inoue,~K.; Kishine,~J. Symmetry, Structure, and
  Dynamics of Monoaxial Chiral Magnets. \emph{J. Phys. Soc. Jpn.}
  \textbf{2016}, \emph{85}, 112001\relax
\mciteBstWouldAddEndPuncttrue
\mciteSetBstMidEndSepPunct{\mcitedefaultmidpunct}
{\mcitedefaultendpunct}{\mcitedefaultseppunct}\relax
\EndOfBibitem
\bibitem[Fert \latin{et~al.}(2017)Fert, Reyren, and
  Cros]{Fert_2017_Nature_rev_skyrmion}
Fert,~A.; Reyren,~N.; Cros,~V. Magnetic skyrmions: advances in physics and
  potential applications. \emph{Nat. Rev. Mater.} \textbf{2017}, \emph{2},
  17031\relax
\mciteBstWouldAddEndPuncttrue
\mciteSetBstMidEndSepPunct{\mcitedefaultmidpunct}
{\mcitedefaultendpunct}{\mcitedefaultseppunct}\relax
\EndOfBibitem
\bibitem[Moriya and Miyadai(1982)Moriya, and Miyadai]{Moriya_SSC_1982}
Moriya,~T.; Miyadai,~T. Evidence for the helical spin structure due to
  antisymmetric exchange interaction in Cr13NbS2. \emph{Solid State Commun.}
  \textbf{1982}, \emph{42}, 209 -- 212\relax
\mciteBstWouldAddEndPuncttrue
\mciteSetBstMidEndSepPunct{\mcitedefaultmidpunct}
{\mcitedefaultendpunct}{\mcitedefaultseppunct}\relax
\EndOfBibitem
\bibitem[Dzyaloshinsky(1958)]{dzyaloshinsky1958}
Dzyaloshinsky,~I. A thermodynamic theory of “weak” ferromagnetism of
  antiferromagnetics. \emph{J. Phys. Chem. Solids} \textbf{1958}, \emph{4}, 241
  -- 255\relax
\mciteBstWouldAddEndPuncttrue
\mciteSetBstMidEndSepPunct{\mcitedefaultmidpunct}
{\mcitedefaultendpunct}{\mcitedefaultseppunct}\relax
\EndOfBibitem
\bibitem[Moriya(1960)]{moriya1960}
Moriya,~T. Anisotropic Superexchange Interaction and Weak Ferromagnetism.
  \emph{Phys. Rev.} \textbf{1960}, \emph{120}, 91--98\relax
\mciteBstWouldAddEndPuncttrue
\mciteSetBstMidEndSepPunct{\mcitedefaultmidpunct}
{\mcitedefaultendpunct}{\mcitedefaultseppunct}\relax
\EndOfBibitem
\bibitem[Miyadai \latin{et~al.}(1983)Miyadai, Kikuchi, Kondo, Sakka, Arai, and
  Ishikawa]{Miyadai_JPSJ_1983}
Miyadai,~T.; Kikuchi,~K.; Kondo,~H.; Sakka,~S.; Arai,~M.; Ishikawa,~Y. Magnetic
  Properties of Cr1\/3NbS2. \emph{J. Phys. Soc. Jpn.} \textbf{1983}, \emph{52},
  1394--1401\relax
\mciteBstWouldAddEndPuncttrue
\mciteSetBstMidEndSepPunct{\mcitedefaultmidpunct}
{\mcitedefaultendpunct}{\mcitedefaultseppunct}\relax
\EndOfBibitem
\bibitem[Shinozaki \latin{et~al.}(2016)Shinozaki, Hoshino, Masaki, Kishine, and
  Kato]{Shinozaki_jpsj_2016}
Shinozaki,~M.; Hoshino,~S.; Masaki,~Y.; Kishine,~J.; Kato,~Y.
  Finite-Temperature Properties of Three-Dimensional Chiral Helimagnets.
  \emph{J. Phys. Soc. Jpn} \textbf{2016}, \emph{85}, 074710\relax
\mciteBstWouldAddEndPuncttrue
\mciteSetBstMidEndSepPunct{\mcitedefaultmidpunct}
{\mcitedefaultendpunct}{\mcitedefaultseppunct}\relax
\EndOfBibitem
\bibitem[Laliena \latin{et~al.}(2016)Laliena, Campo, and
  Kousaka]{Laliena_PRB_16_underHT}
Laliena,~V.; Campo,~J.; Kousaka,~Y. Understanding the H-T phase diagram of the
  monoaxial helimagnet. \emph{Phys. Rev. B} \textbf{2016}, \emph{94},
  094439\relax
\mciteBstWouldAddEndPuncttrue
\mciteSetBstMidEndSepPunct{\mcitedefaultmidpunct}
{\mcitedefaultendpunct}{\mcitedefaultseppunct}\relax
\EndOfBibitem
\bibitem[Yonemura \latin{et~al.}(2017)Yonemura, Shimamoto, Kida, Yoshizawa,
  Kousaka, Nishihara, Goncalves, Akimitsu, Inoue, Hagiwara, and
  Togawa]{Yonemura2017_PRB_phase_diac_oblique}
Yonemura,~J.-i.; Shimamoto,~Y.; Kida,~T.; Yoshizawa,~D.; Kousaka,~Y.;
  Nishihara,~S.; Goncalves,~F. J.~T.; Akimitsu,~J.; Inoue,~K.; Hagiwara,~M.;
  Togawa,~Y. Magnetic solitons and magnetic phase diagram of the hexagonal
  chiral crystal ${\mathrm{CrNb}}_{3}{\mathrm{S}}_{6}$ in oblique magnetic
  fields. \emph{Phys. Rev. B} \textbf{2017}, \emph{96}, 184423\relax
\mciteBstWouldAddEndPuncttrue
\mciteSetBstMidEndSepPunct{\mcitedefaultmidpunct}
{\mcitedefaultendpunct}{\mcitedefaultseppunct}\relax
\EndOfBibitem
\bibitem[Togawa \latin{et~al.}(2015)Togawa, Koyama, Nishimori, Matsumoto,
  McVitie, McGrouther, Stamps, Kousaka, Akimitsu, Nishihara, Inoue, Bostrem,
  Sinitsyn, Ovchinnikov, and Kishine]{Yoshi_PRB_15_confinement}
Togawa,~Y.; Koyama,~T.; Nishimori,~Y.; Matsumoto,~Y.; McVitie,~S.;
  McGrouther,~D.; Stamps,~R.~L.; Kousaka,~Y.; Akimitsu,~J.; Nishihara,~S.;
  Inoue,~K.; Bostrem,~I.~G.; Sinitsyn,~V.~E.; Ovchinnikov,~A.~S.; Kishine,~J.
  Magnetic soliton confinement and discretization effects arising from
  macroscopic coherence in a chiral spin soliton lattice. \emph{Phys. Rev. B}
  \textbf{2015}, \emph{92}, 220412\relax
\mciteBstWouldAddEndPuncttrue
\mciteSetBstMidEndSepPunct{\mcitedefaultmidpunct}
{\mcitedefaultendpunct}{\mcitedefaultseppunct}\relax
\EndOfBibitem
\bibitem[Goncalves \latin{et~al.}(2017)Goncalves, Sogo, Shimamoto, Kousaka,
  Akimitsu, Nishihara, Inoue, Yoshizawa, Hagiwara, Mito, Stamps, Bostrem,
  Sinitsyn, Ovchinnikov, Kishine, and Togawa]{Francisco_PRB_17_FMR}
Goncalves,~F. J.~T. \latin{et~al.}  Collective resonant dynamics of the chiral
  spin soliton lattice in a monoaxial chiral magnetic crystal. \emph{Phys. Rev.
  B} \textbf{2017}, \emph{95}, 104415\relax
\mciteBstWouldAddEndPuncttrue
\mciteSetBstMidEndSepPunct{\mcitedefaultmidpunct}
{\mcitedefaultendpunct}{\mcitedefaultseppunct}\relax
\EndOfBibitem
\bibitem[Mito \latin{et~al.}(2018)Mito, Ohsumi, Tsuruta, Kotani, Nakamura,
  Togawa, Shinozaki, Kato, Kishine, Ohe, Kousaka, Akimitsu, and
  Inoue]{Mito_PRB_18_geometrical_protection}
Mito,~M.; Ohsumi,~H.; Tsuruta,~K.; Kotani,~Y.; Nakamura,~T.; Togawa,~Y.;
  Shinozaki,~M.; Kato,~Y.; Kishine,~J.; Ohe,~J.-i.; Kousaka,~Y.; Akimitsu,~J.;
  Inoue,~K. Geometrical protection of topological magnetic solitons in
  microprocessed chiral magnets. \emph{Phys. Rev. B} \textbf{2018}, \emph{97},
  024408\relax
\mciteBstWouldAddEndPuncttrue
\mciteSetBstMidEndSepPunct{\mcitedefaultmidpunct}
{\mcitedefaultendpunct}{\mcitedefaultseppunct}\relax
\EndOfBibitem
\bibitem[Beleggia \latin{et~al.}(2003)Beleggia, Schofield, Zhu, Malac, Liu, and
  Freeman]{Beleggia_apl_2003_mag_sim}
Beleggia,~M.; Schofield,~M.~A.; Zhu,~Y.; Malac,~M.; Liu,~Z.; Freeman,~M.
  Quantitative study of magnetic field distribution by electron holography and
  micromagnetic simulations. \emph{Appl. Phys. Lett.} \textbf{2003}, \emph{83},
  1435--1437\relax
\mciteBstWouldAddEndPuncttrue
\mciteSetBstMidEndSepPunct{\mcitedefaultmidpunct}
{\mcitedefaultendpunct}{\mcitedefaultseppunct}\relax
\EndOfBibitem
\bibitem[McVitie \latin{et~al.}(2001)McVitie, White, Scott, Warin, and
  Chapman]{Stephen_JAP_2001}
McVitie,~S.; White,~G.~S.; Scott,~J.; Warin,~P.; Chapman,~J.~N. Quantitative
  imaging of magnetic domain walls in thin films using Lorentz and magnetic
  force microscopies. \emph{J. Appl. Phys.} \textbf{2001}, \emph{90},
  5220--5227\relax
\mciteBstWouldAddEndPuncttrue
\mciteSetBstMidEndSepPunct{\mcitedefaultmidpunct}
{\mcitedefaultendpunct}{\mcitedefaultseppunct}\relax
\EndOfBibitem
\bibitem[Kishine and Ovchinnikov(2009)Kishine, and
  Ovchinnikov]{Kishine_PRB_2009_spin_resonance}
Kishine,~J.; Ovchinnikov,~A.~S. Theory of spin resonance in a chiral
  helimagnet. \emph{Phys. Rev. B} \textbf{2009}, \emph{79}, 220405\relax
\mciteBstWouldAddEndPuncttrue
\mciteSetBstMidEndSepPunct{\mcitedefaultmidpunct}
{\mcitedefaultendpunct}{\mcitedefaultseppunct}\relax
\EndOfBibitem
\bibitem[Chapman and Scheinfein(1999)Chapman, and Scheinfein]{CHAPMAN1999729}
Chapman,~J.; Scheinfein,~M. Transmission electron microscopies of magnetic
  microstructures. \emph{J. Magn. Magn. Mater.} \textbf{1999}, \emph{200}, 729
  -- 740\relax
\mciteBstWouldAddEndPuncttrue
\mciteSetBstMidEndSepPunct{\mcitedefaultmidpunct}
{\mcitedefaultendpunct}{\mcitedefaultseppunct}\relax
\EndOfBibitem
\bibitem[Kishine \latin{et~al.}(2005)Kishine, Inoue, and
  Yoshida]{Kishine_PTPS_05}
Kishine,~J.; Inoue,~K.; Yoshida,~Y. Synthesis, Structure and Magnetic
  Properties of Chiral Molecule-Based Magnets). \emph{Progr. Theor. Phys.
  Suppl.} \textbf{2005}, \emph{159}, 82--95\relax
\mciteBstWouldAddEndPuncttrue
\mciteSetBstMidEndSepPunct{\mcitedefaultmidpunct}
{\mcitedefaultendpunct}{\mcitedefaultseppunct}\relax
\EndOfBibitem
\bibitem[{Dussaux A.} \latin{et~al.}(2016){Dussaux A.}, {Schoenherr P.},
  {Koumpouras K.}, {Chico J.}, {Chang K.}, {Lorenzelli L.}, {Kanazawa N.},
  {Tokura Y.}, {Garst M.}, {Bergman A.}, {Degen C. L.}, and {Meier
  D.}]{dussaux_ncoms_16_FeGe_dynamics}
{Dussaux A.},; {Schoenherr P.},; {Koumpouras K.},; {Chico J.},; {Chang K.},;
  {Lorenzelli L.},; {Kanazawa N.},; {Tokura Y.},; {Garst M.},; {Bergman A.},;
  {Degen C. L.},; {Meier D.}, {Local dynamics of topological magnetic defects
  in the itinerant helimagnet FeGe}. \emph{Nat. Commun.} \textbf{2016},
  \emph{7}, 12430\relax
\mciteBstWouldAddEndPuncttrue
\mciteSetBstMidEndSepPunct{\mcitedefaultmidpunct}
{\mcitedefaultendpunct}{\mcitedefaultseppunct}\relax
\EndOfBibitem
\bibitem[Shinozaki \latin{et~al.}(2018)Shinozaki, Masaki, Aoki, Togawa, and
  Kato]{Shinozaki_PRB18_soliton_barrier}
Shinozaki,~M.; Masaki,~Y.; Aoki,~R.; Togawa,~Y.; Kato,~Y. Intrinsic hysteresis
  due to the surface barrier for chiral solitons in monoaxial chiral
  helimagnets. \emph{Phys. Rev. B} \textbf{2018}, \emph{97}, 214413\relax
\mciteBstWouldAddEndPuncttrue
\mciteSetBstMidEndSepPunct{\mcitedefaultmidpunct}
{\mcitedefaultendpunct}{\mcitedefaultseppunct}\relax
\EndOfBibitem
\bibitem[Goncalves \latin{et~al.}(2018)Goncalves, Sogo, Shimamoto, Proskurin,
  Sinitsyn, Kousaka, Bostrem, Kishine, Ovchinnikov, and
  Togawa]{FTG_PRB18_tailored_FMR}
Goncalves,~F. J.~T.; Sogo,~T.; Shimamoto,~Y.; Proskurin,~I.; Sinitsyn,~V.~E.;
  Kousaka,~Y.; Bostrem,~I.~G.; Kishine,~J.; Ovchinnikov,~A.~S.; Togawa,~Y.
  Tailored resonance in micrometer-sized monoaxial chiral helimagnets.
  \emph{Phys. Rev. B} \textbf{2018}, \emph{98}, 144407\relax
\mciteBstWouldAddEndPuncttrue
\mciteSetBstMidEndSepPunct{\mcitedefaultmidpunct}
{\mcitedefaultendpunct}{\mcitedefaultseppunct}\relax
\EndOfBibitem
\bibitem[Bar'yakhtar \latin{et~al.}(1968)Bar'yakhtar, Savchenco, and
  Tarasenko]{Baryakhtar_JETP_68_FM_dislocation}
Bar'yakhtar,~V.~G.; Savchenco,~M.~A.; Tarasenko,~V.~V. Effect of Dislocations
  on the Line Width of Uniform Ferro- and Antiferromagnetic Resonances.
  \emph{J. Exp. Theor. Phys.} \textbf{1968}, \emph{25}, 858--862\relax
\mciteBstWouldAddEndPuncttrue
\mciteSetBstMidEndSepPunct{\mcitedefaultmidpunct}
{\mcitedefaultendpunct}{\mcitedefaultseppunct}\relax
\EndOfBibitem
\bibitem[Kittel(1948)]{kittel1948}
Kittel,~C. On the Theory of Ferromagnetic Resonance Absorption. \emph{Phys.
  Rev.} \textbf{1948}, \emph{73}, 155--161\relax
\mciteBstWouldAddEndPuncttrue
\mciteSetBstMidEndSepPunct{\mcitedefaultmidpunct}
{\mcitedefaultendpunct}{\mcitedefaultseppunct}\relax
\EndOfBibitem
\bibitem[Togawa \latin{et~al.}(2013)Togawa, Kousaka, Nishihara, Inoue,
  Akimitsu, Ovchinnikov, and Kishine]{togawa_2013_PRL_MR}
Togawa,~Y.; Kousaka,~Y.; Nishihara,~S.; Inoue,~K.; Akimitsu,~J.;
  Ovchinnikov,~A.~S.; Kishine,~J. Interlayer Magnetoresistance due to Chiral
  Soliton Lattice Formation in Hexagonal Chiral Magnet
  ${\mathrm{CrNb}}_{3}{\mathrm{S}}_{6}$. \emph{Phys. Rev. Lett.} \textbf{2013},
  \emph{111}, 197204\relax
\mciteBstWouldAddEndPuncttrue
\mciteSetBstMidEndSepPunct{\mcitedefaultmidpunct}
{\mcitedefaultendpunct}{\mcitedefaultseppunct}\relax
\EndOfBibitem
\bibitem[Kishine \latin{et~al.}(2016)Kishine, Proskurin, Bostrem, Ovchinnikov,
  and Sinitsyn]{Kishine_PRB_16_pined_CSL_resonance}
Kishine,~J.; Proskurin,~I.; Bostrem,~I.~G.; Ovchinnikov,~A.~S.; Sinitsyn,~V.~E.
  Resonant collective dynamics of the weakly pinned soliton lattice in a
  monoaxial chiral helimagnet. \emph{Phys. Rev. B} \textbf{2016}, \emph{93},
  054403\relax
\mciteBstWouldAddEndPuncttrue
\mciteSetBstMidEndSepPunct{\mcitedefaultmidpunct}
{\mcitedefaultendpunct}{\mcitedefaultseppunct}\relax
\EndOfBibitem
\bibitem[McVitie \latin{et~al.}(2015)McVitie, McGrouther, McFadzean, MacLaren,
  O’Shea, and Benitez]{mcvitie2015_magtem}
McVitie,~S.; McGrouther,~D.; McFadzean,~S.; MacLaren,~D.; O’Shea,~K.;
  Benitez,~M. Aberration corrected Lorentz scanning transmission electron
  microscopy. \emph{Ultramicroscopy} \textbf{2015}, \emph{152}, 57 -- 62\relax
\mciteBstWouldAddEndPuncttrue
\mciteSetBstMidEndSepPunct{\mcitedefaultmidpunct}
{\mcitedefaultendpunct}{\mcitedefaultseppunct}\relax
\EndOfBibitem
\bibitem[Chapman \latin{et~al.}(1990)Chapman, McFadyen, and
  McVitie]{chapman_1990_mdpc}
Chapman,~J.~N.; McFadyen,~I.~R.; McVitie,~S. Modified differential phase
  contrast Lorentz microscopy for improved imaging of magnetic structures.
  \emph{IEEE Trans. Magn.} \textbf{1990}, \emph{26}, 1506--1511\relax
\mciteBstWouldAddEndPuncttrue
\mciteSetBstMidEndSepPunct{\mcitedefaultmidpunct}
{\mcitedefaultendpunct}{\mcitedefaultseppunct}\relax
\EndOfBibitem
\bibitem[Malis \latin{et~al.}(1988)Malis, Cheng, and Egerton]{Malis1988_tol}
Malis,~T.; Cheng,~S.~C.; Egerton,~R.~F. EELS log-ratio technique for
  specimen-thickness measurement in the TEM. \emph{J. Elec. Microsc. Tech.}
  \textbf{1988}, \emph{8}, 193--200\relax
\mciteBstWouldAddEndPuncttrue
\mciteSetBstMidEndSepPunct{\mcitedefaultmidpunct}
{\mcitedefaultendpunct}{\mcitedefaultseppunct}\relax
\EndOfBibitem
\bibitem[fpd()]{fpd}
{FPD: Fast pixelated detector data storage, analysis and visualisation.
  https://gitlab.com/fpdpy/fpd (accessed January 17, 2019)}\relax
\mciteBstWouldAddEndPuncttrue
\mciteSetBstMidEndSepPunct{\mcitedefaultmidpunct}
{\mcitedefaultendpunct}{\mcitedefaultseppunct}\relax
\EndOfBibitem
\bibitem[de~la Peña \latin{et~al.}(2018)de~la Peña, Fauske, Burdet, Prestat,
  Jokubauskas, Nord, Ostasevicius, MacArthur, Sarahan, Johnstone, Taillon,
  Eljarrat, Migunov, Caron, Furnival, Mazzucco, Aarholt, Walls, Slater,
  Winkler, Martineau, Donval, McLeod, Hoglund, Alxneit, Hjorth, Henninen,
  Zagonel, Garmannslund, and Skorikov]{hyperspy}
de~la Peña,~F. \latin{et~al.}  hyperspy/hyperspy v1.4.1. 2018;
  \url{https://doi.org/10.5281/zenodo.1469364}\relax
\mciteBstWouldAddEndPuncttrue
\mciteSetBstMidEndSepPunct{\mcitedefaultmidpunct}
{\mcitedefaultendpunct}{\mcitedefaultseppunct}\relax
\EndOfBibitem
\bibitem[{Fujiyoshi Yoshinori}(2011)]{Yoshinori_2011_cryo_tem}
{Fujiyoshi Yoshinori}, {Structural physiology based on electron
  crystallography}. \emph{Protein science : a publication of the Protein
  Society} \textbf{2011}, \emph{20}, 806–817\relax
\mciteBstWouldAddEndPuncttrue
\mciteSetBstMidEndSepPunct{\mcitedefaultmidpunct}
{\mcitedefaultendpunct}{\mcitedefaultseppunct}\relax
\EndOfBibitem
\bibitem[Abbaschian \latin{et~al.}(2008)Abbaschian, Abbaschian, and
  Reed-Hill]{Physical_Metallurgy_Principles}
Abbaschian,~R.; Abbaschian,~L.; Reed-Hill,~R.~E. \emph{Physical Metallurgy
  Principles}; Cengage Learning, 2008\relax
\mciteBstWouldAddEndPuncttrue
\mciteSetBstMidEndSepPunct{\mcitedefaultmidpunct}
{\mcitedefaultendpunct}{\mcitedefaultseppunct}\relax
\EndOfBibitem
\bibitem[Li \latin{et~al.}(2017)Li, Zheng, Tavabi, Caron, Jin, Du, Kovács,
  Tian, Farle, and Dunin-Borkowski]{li_2017_FeGe_dislocation_crystal}
Li,~Z.-A.; Zheng,~F.; Tavabi,~A.~H.; Caron,~J.; Jin,~C.; Du,~H.; Kovács,~A.;
  Tian,~M.; Farle,~M.; Dunin-Borkowski,~R.~E. Magnetic Skyrmion Formation at
  Lattice Defects and Grain Boundaries Studied by Quantitative Off-Axis
  Electron Holography. \emph{Nano Lett.} \textbf{2017}, \emph{17},
  1395--1401\relax
\mciteBstWouldAddEndPuncttrue
\mciteSetBstMidEndSepPunct{\mcitedefaultmidpunct}
{\mcitedefaultendpunct}{\mcitedefaultseppunct}\relax
\EndOfBibitem
\end{mcitethebibliography}

\providecommand{\latin}[1]{#1}
\providecommand*\mcitethebibliography{\thebibliography}
\csname @ifundefined\endcsname{endmcitethebibliography}
  {\let\endmcitethebibliography\endthebibliography}{}

\end{document}